\documentclass{aa}
\usepackage[varg]{txfonts}
\usepackage{natbib}
\bibpunct{(}{)}{;}{a}{}{,}
\usepackage[
  colorlinks=true,
  linkcolor=blue,
  citecolor=blue,
  urlcolor=blue
]{hyperref}
\usepackage{xspace}
\usepackage{amssymb}
\usepackage{amsmath}
\usepackage{graphicx}
\usepackage{placeins}

\defcitealias{2019MNRAS.484..728M}{Mc19}
\defcitealias{2020MNRAS.493.5892W}{WB20}

\begin{document}

\title{Formation of multiple dust rings and gaps in protoplanetary discs by a single migrating planet}
\subtitle{A parameter study in locally isothermal discs}
\titlerunning{Parameter study in locally isothermal discs}
\authorrunning{S.~C.~Meiners et al.}
\author{S.~C.~Meiners\inst{\ref{ZAH}}\and K.~M.~Weiskopf\inst{\ref{LMU}}\and T.~Rometsch\inst{\ref{ZAH}}\and C.~P.~Dullemond\inst{\ref{ZAH}}}
\institute{
  Institute for Theoretical Astrophysics, Centre for Astronomie (ZAH)\label{ZAH}, Heidelberg University, Albert Ueberle Str. 2, 69120 Heidelberg, Germany 
  \and Ludwig-Maximilians-Universität München, Universitäts-Sternwarte, Scheinerstr. 1, 81679 München, Germany\label{LMU}
  }
\date{Received date /
Accepted date }

\abstract
{ALMA observations reveal that large protoplanetary discs usually contain multiple concentric dust rings separated by dark gaps. A natural explanation is dust-trapping at the edges of gaps opened by newly formed planets. However, planets are known to migrate inward on timescales shorter than the typical disc lifetime, seemingly at odds with observations of rings at large radii.}
{We wish to investigate the conditions under which migrating planets can form long-lived, multi-ringed structures out to $r\sim 150\,\mathrm{au}$. Thus, we intend to constrain the viable parameter space for the planetary origin hypothesis of the rings.}
{We used the \texttt{FargoCPT} hydrodynamics code and set up two-dimensional models of locally isothermal protoplanetary discs with a single migrating planet. We varied key parameters such as the disc aspect ratio (temperature), viscous parameter ($\alpha$), and planetary mass.}
{In all our models, the planet eventually stalls in a deep gap. At $\alpha\leq 10^{-4}$, secondary spirals launched by the planet have the capacity to open additional gaps at smaller radii. When planets exceed twice the local thermal mass prior to the stall, they enter a regime of alternating slow and type-III rapid migration, which leaves partial gaps outside the planet's orbit. The gap edges consistently feature pressure maxima capable of dust trapping. While these start out as large-scale vortices at $\alpha\leq 10^{-4}$, they gradually smear out into ring-like structures prior to their dissipation. The 'remnant' rings left by type-III migration quickly dissipate at $\alpha=10^{-3}$, but persist for at least $300\text{--}500\,\mathrm{kyr}$ at $\alpha\leq 10^{-4}$. Both the smear-out and dissipation timescales increase with lower $\alpha$.}
{Our results indicate that migrating planets can reproduce the observed multi-ringed structures in discs with $\alpha\lesssim 10^{-4}$ through their stall ($r\lesssim 50\,\mathrm{au}$), secondary gap-opening ($r\lesssim 20\,\mathrm{au}$), and type-III migration remnants (extending to $r\sim 150\,\mathrm{au}$ for Jupiter-mass planets in sufficiently massive discs). Longer simulations will be required to compare the statistics of ring-to-vortex occurrence to observations.}


\maketitle
\nolinenumbers

\section{Introduction}
\label{sec:introduction}
One of the most striking features of ALMA observations is that most large protoplanetary discs consist of concentric dust rings separated by gaps \citep{2015ApJ...808L...3A,2018ApJ...869L..41A}. The dust in these rings is likely trapped at a so-called pressure bump (pressure maxima) in the underlying gas distribution of the disc. From a theoretical perspective, the existence of such ring-shaped dust traps has long been suspected \citep{1972fpp..conf..211W,2007ApJ...664L..55K}. It was proposed by \citet{2012A&A...538A.114P} as an answer to the mystery of why old protoplanetary discs still contain large amounts of dust in their outer regions \citep{2007A&A...469.1169B}. The multi-ringed appearance of protoplanetary discs subsequently observed with ALMA therefore seems to confirm this picture. It is now even possible to measure the efficiency of dust-trapping in a given dust ring by mapping the deviation from Keplerian rotation of the gas across these pressure bumps with CO channel maps \citep{2020MNRAS.495..173R}.

While this picture is appealing and accounts for many features of the ALMA continuum maps, it rests on the assumption that variations in the radial gas surface density $\Sigma_{\mathrm{g}}(r)$ are strong enough to produce midplane pressure maxima capable of dust-trapping. The origin of these variations, however, remains unclear. One possible explanation is that they could be a relic from the disc's creation. Yet if protoplanetary discs are turbulently viscous \citep{2012ARA&A..50..211K}, then any such early variations would quickly smear out. Ultimately, we presume that there must be a physical process that creates and/or maintains these variations in $\Sigma_{\mathrm{g}}(r)$.

Numerous propositions have been made to potentially explain the existence of these variations, including viscosity transitions at dead-zone boundaries \citep{2001MNRAS.324..705A,2006A&A...446L..13V,2015A&A...574A..68F,2017ApJ...835..230F} or at snowlines \citep{2007ApJ...664L..55K,2013ApJ...765..114D,2014A&A...570A..75B}, as well as magnetically driven zonal flows \citep{2009ApJ...697.1269J,2012MNRAS.422.2685S,2014ApJ...796...31B,2015ApJ...798...84B}. Gravitational processes have also been proposed, ranging from classical gravitational instability in sufficiently massive discs \citep{2004MNRAS.355..543R,2007prpl.conf..607D} to the secular gravitational instability, which can occur even when the disc is Toomre-stable \citep{2014ApJ...794...55T}.

Another commonly proposed explanation would be that the gaps between the rings are caused by planets. Recent observations have, in fact, confirmed the presence of planets in the gaps of protoplanetary discs around PDS\,70 \citep{2018A&A...617A..44K, 2019NatAs...3..749H} and WISPIT\,2 \citep{2025ApJ...990L...9C, 2026ApJ..1000L..38L}. These detections align with the theoretical picture, in which a massive enough planet opens a circular gap in the gas disc \citep{2012ARA&A..50..211K}. At the gap edges, a local gas pressure maximum naturally appears, which traps the dust \citep{2012A&A...538A.114P}. This scenario has been tested and appears to explain the dust rings seen in many protoplanetary discs \citep{2018ApJ...869L..47Z, 2021A&A...647A.174R, 2023ApJ...945L..37G, 2023MNRAS.524.3930Z}. Based on the depth and width of the gaps between the dust rings, the masses of the putative planets can be directly estimated \citep{2019MNRAS.486..453L, 2021ApJ...923..165W}, although it was shown that a single planet can also induce multiple gaps and dust traps \citep{2017ApJ...843..127D, 2017ApJ...850..201B, 2018ApJ...869L..47Z}.

In the above studies, the effect of the planet onto the disc is studied under the assumption that the planet's orbit is kept fixed. However, the same planet--disc interaction that produces the gaps and dust traps also causes the planet to migrate \citep{2012ARA&A..50..211K}. Usually this migration is directed inward, and can occur on timescales smaller than the typical disc lifetime of $\sim3\text{--}5\,\mathrm{Myr}$ \citep{armitage2020}. At first glance, this appears difficult to reconcile with a planetary origin for dust rings observed at large stellocentric radii, out to $r\sim150\,\mathrm{au}$ \citep{2018ApJ...869L..42H}, which would seem to require the planets themselves to remain at such distances.

This apparent contradiction may be resolved by the 'intermittent migration' regime, recently identified by \citet{2019MNRAS.484..728M} (hereafter, \citetalias{2019MNRAS.484..728M}) and \citet{2020MNRAS.493.5892W} (hereafter, \citetalias{2020MNRAS.493.5892W}) through studies of single-planet migration in locally isothermal discs. They found that partial gap-opening planets can alternate between episodes of slow and rapid migration. During the slow migration episodes, the planet forms ring and gap structures, which are then left behind in subsequent rapid migration episodes. In this way, a single planet can generate multiple rings and gaps out to large stellocentric distances despite undergoing globally fast inward migration.

However, these remnant structures are no longer directly supported by the planet once it has migrated away. As a result, they are subject to viscous smoothing, causing them to gradually diffuse and eventually disappear entirely. While several recent studies suggest that discs may be less viscous than previously assumed, with a turbulent $\alpha$-parameter \citep{1973A&A....24..337S} of around $\alpha\sim10^{-4}$ \citep{2016ApJ...816...25P,2018ApJ...869L..46D,2022ApJ...930...11V,2023NewAR..9601674R}, enabling the long-term survival of left-behind structures, low viscosity also favours the formation and survival of vortices \citep{2007A&A...471.1043D}. This stands in contrast to the prevalence of axisymmetric ring structures in a majority of protoplanetary discs, although some vortices have also been observed in ALMA maps \citep{2018ApJ...869L..41A}. Thus, if intermittent migration is invoked to address the issue of ring and gap formation by migrating planets, viscosity needs to be low enough to ensure long lifetimes of the structures left behind, but high enough that vortices are either absent or short-lived. 

The purpose of our paper is to follow up on the studies of \citetalias{2019MNRAS.484..728M} and \citetalias{2020MNRAS.493.5892W} with long runs at high spatial resolution and a wide parameter scan tailored to ALMA observations. We aim to find out whether migrating planets can form multiple ring structures up to large radii as observed in many protoplanetary discs, and, in particular, whether this can occur through the intermittent migration regime. We seek to analyse the lifetime of structures and statistics of ring-to-vortex occurrence. If possible, we intend to use these results to set constraints on some physical properties of the discs.

\section{Theory}
\label{sec:theory}
As previously mentioned, the interaction between a migrating planet and its surrounding protoplanetary disc is inherently complex and highly sensitive to variations across numerous physical parameters. Over recent decades, extensive theoretical and numerical studies have sought to clarify the underlying mechanisms and to classify the resulting migration regimes. In the following, we provide a concise overview of the theoretical framework necessary to interpret the diverse migration behaviours and disc structures that arise in our simulations.

An embedded planet interacts gravitationally with the surrounding disc, leading to the excitation of spiral wakes at specific resonant locations known as Lindblad resonances \citep{1980ApJ...241..425G}, as well as to the establishment of a corotation region around the planet's orbit, where gas executes horseshoe orbits relative to the planet \citep{1991LPI....22.1463W}. Both features exert torques on the planet \citep{2012ARA&A..50..211K}, with a net negative torque inducing inward migration and a net positive torque inducing outward migration. The outer Lindblad spiral and gas on outward (trailing) horseshoe turns exert a negative torque, while the inner Lindblad spiral and gas on inward (leading) horseshoe turns exert a positive torque.

At a certain radial distance from the planet, the Lindblad spirals, propagating with a pattern speed equal to the planet's orbital frequency, become transonic with respect to the local gas flow, turning into shock waves \citep{2001ApJ...552..793G, 2002ApJ...569..997R}. As these shocks dissipate, positive angular momentum is deposited in the outer disc, while negative angular momentum is deposited in the inner disc, causing the disc matter to recede from the planet \citep{2001ApJ...552..793G}. Provided that this tidal force can overcome the viscous and pressure forces working against it, an annular gap opens in the gas surface density around the planet's orbit. As the corotation region and eventually the Lindblad resonant locations are drained of material due to the gap opening, the corresponding torques weaken, substantially slowing down the planet's migration \citep{2002ApJ...569..997R}.

Evidently, the migration behaviour of an embedded planet is closely linked to its ability to open a gap in the local gas surface density. In low-viscosity discs, two critical planet mass thresholds for gap opening can be defined: The first is the thermal mass $M_{\mathrm{th}}$, above which the planet's wake shocks immediately upon excitation at the Lindblad resonances \citep{2001ApJ...552..793G}, given by
\begin{equation}
    \label{eq:thermalmass}
    M_{\mathrm{th}} = \frac{2}{3}h^{3}M_{\star},
\end{equation}
where $h$ is the local disc aspect ratio (see Sect.~\ref{sec:setup}) and $M_{\star}$ is the mass of the central star. The second is the feedback mass $M_{\mathrm{f}}$, above which the planet starts to significantly alter the local gas surface density, causing 'feedback' effects on its migration \citep{2002ApJ...572..566R}, given by
\begin{equation}
    \label{eq:feedbackmass}
    M_{\mathrm{f}} \approx 3.8 \left(\frac{Q}{h}\right)^{-5/13}M_{\mathrm{th}},
\end{equation}
with the Toomre parameter \citep{1964ApJ...139.1217T}, which indicates that the disc is gravitationally stable for
\begin{equation}
    \label{eq:toomre}
    Q=\frac{c_{\mathrm{s}}\Omega_{\mathrm{K}}}{\pi G\Sigma} \gg 1,
\end{equation}
where $c_{\mathrm{s}}$ is the isothermal sound speed, $\Omega_{\mathrm{K}}$ is the Keplerian angular frequency, $G$ is the gravitational constant, and $\Sigma$ is the gas surface density. While neither of these mass thresholds should be considered a sharp criterion, they are nonetheless useful in the classification of planetary migration regimes. Both \citetalias{2019MNRAS.484..728M} and \citet{2025MNRAS.542.1685Z}, whom we refer to for a detailed mapping of migration regimes (see Fig.~1 in \citetalias{2019MNRAS.484..728M} and Fig.~18 in \citealt{2025MNRAS.542.1685Z}), defined the three primary regimes in low-viscosity discs accordingly:

\textbf{Type-I:} While planets with a mass of $M_{\mathrm{p}} \lesssim M_{\mathrm{f}}$ can excite Lindblad spirals in the disc and form a corotation region around their orbit, they are not massive enough to open a significant gap. Consequently, their migration is governed by the Lindblad and corotation torques \citep{1980ApJ...241..425G, 2012ARA&A..50..211K}. Migration is typically driven by a dominant negative Lindblad torque that is moderated by a positive corotation torque. This results in a fast inward migration, with a migration timescale of $\tau_{mig}\sim10^5\,\mathrm{yr}$ for an Earth-mass planet \citep{2012ARA&A..50..211K}.

\textbf{Feedback:} Planets with a mass of $M_{\mathrm{f}} \lesssim M_{\mathrm{p}} \lesssim M_{\mathrm{th}}$ are able to open a partial gap, leading to a moderate reduction of the Lindblad and corotation torques. For a migrating planet, the emerging gap structure is asymmetric \citep{1989ApJ...347..490W, 2025MNRAS.542.1685Z}, with a deeper gap trailing the planet outside its orbit in the case of inward migration. Consequently, the outer Lindblad torque driving the inward migration is more substantially reduced than the inner, leading to a decrease in migration rate until the planet eventually stalls. \citetalias{2019MNRAS.484..728M} referred to this as the 'smooth feedback' regime, where disc viscosity is too low to close the gap but sufficient to prevent vortex formation, usually corresponding to $\alpha\sim10^{-4}$ \citep{2009ApJ...690L..52L, 2010ApJ...712..198Y}.

However, in lower viscosity discs, small-scale, short-lived vortices can form at the gap edges due to the Rossby wave instability (RWI; \citealt{1999ApJ...513..805L, 2007A&A...471.1043D}). These vortices can maintain the Lindblad torque by diffusively refilling the gap, allowing for episodes of sustained faster migration, referred to as vortex-assisted migration (\citetalias{2019MNRAS.484..728M}; \citealt{2025MNRAS.542.1685Z}).

In sufficiently massive discs, planets in the feedback regime can also experience episodes of type-III runaway migration. This runaway is driven by a strong dynamical corotation torque: the migration rate grows high enough that material freshly entering the corotation region crosses in a single horseshoe turn without being trapped in libration, resulting in a self-reinforcing increase in migration rate \citep{2003ApJ...588..494M}. The runaway is moderated both by the reservoir of material available to drive migration and by the added inertia of material still trapped in libration with the planet. \citet{2003ApJ...588..494M} found that this type-III runaway then occurs for $\delta m > M_{\mathrm{p}}$, where $\delta m$ is the coorbital vorticity-weighted mass deficit, expressed as
\begin{equation}
    \label{eq:massdeficit}
    \delta m = 4\pi r_\mathrm{p} \omega(r_\mathrm{p}) \times \left[x_s\mathcal{IV}\left(r_\mathrm{p}-x_s\right)-\int_{r_\mathrm{p}-x_s}^{r_\mathrm{p}}\mathcal{IV}(r)dr\right],
\end{equation}
with the radial distance between star and planet, $r_{\mathrm{p}}$, the radial half-width of the planet's corotation region, $x_{\mathrm{s}}$, the gas vorticity, $\omega=\boldsymbol{\nabla}\times\boldsymbol{u}$, and the inverse vortensity, $\mathcal{IV}=\Sigma/\omega$. The first term in brackets concerns material entering the corotation region at $r_{\mathrm{p}}-x_{\mathrm{s}}$, the second material trapped in libration in the region. \citetalias{2019MNRAS.484..728M} and \citetalias{2020MNRAS.493.5892W} found that the rapid episodes of intermittent migration are instances of type-III runaways. They are fed by a reservoir of material built up at the edge of the partial gap during the preceding slow migration episode. Once this reservoir is exhausted, the runaway stops and slow migration resumes.

\textbf{Type-II:} Planets with a mass of $M_{\mathrm{p}} \gtrsim M_{\mathrm{th}}$ are able to open a deep gap \citep{2006Icar..181..587C}, leading to a substantial reduction of the Lindblad and corotation torques, to the point where migration practically stalls. Continued slow migration, typically directed inward, can be driven by the remaining Lindblad torque (as the planet is unable to deplete the gap entirely of gas; \citealt{2014ApJ...792L..10D}), or by the viscous evolution of the disc \citep{1986ApJ...309..846L}.

Once again, faster episodes of migration due to vortex-assistance are possible. For more massive planets, the gap edge can instead feature a single, large-scale and long-lived vortex, driving migration in a similar manner, in a process referred to as vortex-driven migration \citep{2022A&A...658A..32L}.

As both $M_{\mathrm{th}}$ and $M_{\mathrm{f}}$ are increasing functions of $r$ due to their dependency on $h$, a planet can switch from the type-I regime to the feedback regime or from the feedback regime into the type-II regime as it migrates inward, even if its mass remains constant.

\section{Methods}
\label{sec:methods}
We performed two-dimensional hydrodynamical simulations of vertically integrated protoplanetary discs with an embedded planet. The simulations were carried out using \texttt{FargoCPT} \citep{2024A&A...684A.192R}, an adaptation of the grid-based code \texttt{FARGO} \citep{2000A&AS..141..165M}. \texttt{FargoCPT} utilises finite difference methods to simulate the gas component of the disc and it has the capacity to model the dust component as Lagrangian particles. However, to avoid an additional set of dust parameters and increased computational costs, we conducted our simulations without the dust component. Instead, we relied on the appearance and strength of pressure maxima in the gas distribution to infer the presence and longevity of dust rings.

\subsection{Physical model}
\label{sec:model}
We adopted polar coordinates $(r,\varphi)$ centred on a solar mass star and embedded the planet into the surrounding disc at an initial distance of $r_\mathrm{p0}=r_0 = 50\,\mathrm{au}$. The disc gas was assumed to be ideal with a locally isothermal equation of state, such that the pressure is given by\begin{equation}
    \label{eq:pressure}
    P=c_{\mathrm{s}}^2(r)\Sigma(r),
\end{equation}
with a gas surface density, $\Sigma$, and the isothermal sound speed, $c_{\mathrm{s}}^2=k_\mathrm{B}T/\mu m_\mathrm{p}$, where $k_\mathrm{B}$ is the Boltzmann constant, $\mu=2.35$ is the mean molecular weight, and $m_\mathrm{p}$ is the proton mass. The temperature, $T$, follows a radial power law, which is fixed in our models by our choice of the disc aspect ratio profile, $h$, through the relation $h(r)=H/r=c_{\mathrm{s}}/(\Omega_{\mathrm{K}}r)$, where $H$ is the disc scale height and $\Omega_{\mathrm{K}}$ is the Keplerian angular frequency, expressed as
\begin{equation}
    \label{eq:aspectratio_temperatureprofile}
    h(r) = h_0 \left( \frac{r}{r_0} \right)^{\frac{1}{4}} \Rightarrow T(r) = T_0 \left( \frac{r}{r_0} \right)^{-\frac{1}{2}}.
\end{equation}
Here, $h_0$ and $T_0$ are the aspect ratio and temperature at $r_0$, respectively. The evolution of the vertically integrated gas is then governed by the continuity and Navier-Stokes equations, 
\begin{align}
    \label{eq:continuity}
    &\frac{\partial \Sigma}{\partial t} + \nabla \cdot (\Sigma \boldsymbol{u}) = 0,\\
    \label{eq:navierstokes}
    &\frac{\partial (\Sigma \boldsymbol{u})}{\partial t} + \nabla\cdot\left(\Sigma\boldsymbol{u}\otimes\boldsymbol{u}\right) = - \nabla P - \Sigma \boldsymbol{g} + \nabla \tau,
\end{align}
with the viscous stress tensor, $\tau$, the gas velocity field, $\boldsymbol{u}$, and accelerations due to gravity, $\boldsymbol{g}$. We modelled the radial transport of angular momentum in the disc by a turbulent viscosity, $\nu = \alpha c_{\mathrm{s}} H$, with the dimensionless parameter, $\alpha$ \citep{1973A&A....24..337S}.

Due to the acceleration of the star by the planet and the disc, our stellocentric coordinate frame is non-inertial, introducing additional forces on the planet and the disc. To correct for these forces, we included the so-called indirect terms, $\mathbf{IT}$, alongside the direct acceleration due to the gravitational potential of the star, $\Phi_\star$, and the planet, $\Phi_\mathrm{p}$, 
\begin{equation}
  \label{eq:gravitational_acceleration}
  \boldsymbol{g}=-\nabla(\Phi_\star+\Phi_\mathrm{p})+\mathbf{ITpd}+\mathbf{ITdd}.
\end{equation}
Here, $\mathbf{ITpd}$ is the indirect term correcting for the indirect acceleration of the disc due to the acceleration of the star by the planet, while $\mathbf{ITdd}$ is analogous for the disc on the disc. We further included the indirect term for the disc on the planet ($\mathbf{ITdp}$) and the planet on the planet ($\mathbf{ITpp}$) for the evolution of the $N$-body system. The indirect terms correcting for the acceleration of the star by the planet and the disc are given by
\begin{align}
  \label{eq:indirect_term_planet}
  &\mathbf{ITp}=- GM_\mathrm{p}\frac{\boldsymbol{r_\mathrm{p}}}{\lvert r_\mathrm{p}\rvert^3},\\
  \label{eq:indirect_term_disc}
  &\mathbf{ITd}= - \int_\mathrm{disc} G \Sigma(\boldsymbol{r})\frac{\boldsymbol{r}}{\lVert \boldsymbol{r} \rVert^3} {\rm d}^2 \boldsymbol{r}.
\end{align}
The application of $\mathbf{ITp}$ onto the planet yields $\mathbf{ITpp}$. Applying it onto every disc element yields $\mathbf{ITpd}$. The same holds for $\mathbf{ITd}$ with respect to $\mathbf{ITdp}$ and $\mathbf{ITdd}$.

However, as the Toomre parameter $Q>1$ (see Eq.~\ref{eq:toomre}) over the entire disc, we did not include disc self-gravity. According to \citet{2025OJAp....8E..84C}, this is inconsistent with our application of $\mathbf{ITdd}$, as indirect terms should only be applied when the corresponding direct term is included as well. We address this issue in Sect.~\ref{sec:caveats}.

To account for the vertical extension of the disc, \texttt{FargoCPT} applies a smoothing dependent on the disc scale height, $H$, to $\Phi_\star$ and $\Phi_\mathrm{p}$ \citep{2024A&A...684A.192R}. The gravitational potential at $\boldsymbol{R}$ due to a point mass $i$ with mass, $M_i$, located at $\boldsymbol{R_i}$ is then given by
\begin{equation}
    \label{eq:gravsmoothing}
    \begin{aligned}
    \Phi_i&=-\frac{GM_i}{\sqrt{d^2+s^2}}, & \boldsymbol{d}&=\boldsymbol{R}-\boldsymbol{R_i},
    \end{aligned}
\end{equation}
with a smoothing length of $s=\epsilon H$ and a smoothing parameter of $\epsilon=0.6$ \citep{2012A&A...541A.123M}.

We initialised the gas surface density as 
\begin{equation}
    \label{eq:surfacedensityprofile}
    \Sigma(r) = \Sigma_0 \left( \frac{r}{r_0} \right)^{-1} C_{\mathrm{i}}(r) C_{\mathrm{o}}(r),
\end{equation}
where $\Sigma_0=10$ g cm$^{-2}$ is the initial gas surface density at $r_0$, and $C_{\mathrm{i}}(r)$ and $C_{\mathrm{o}}(r)$ are the inner and outer exponential profile cutoffs, respectively. The inner and outer profile cutoffs are given by
\begin{equation}
\label{eq:cutoff}
    C_{\mathrm{i/o}} = \left(1+\exp\left(\frac{\mp r\pm r_{\mathrm{iC/oC}}}{w_{\mathrm{iC/oC}}}\right)\right)^{-1},
\end{equation}
where $r_{\mathrm{iC/oC}}$ are the cutoff points and $w_{\mathrm{iC/oC}}$ the cutoff widths. We implemented the inner profile cutoff with $r_{\mathrm{iC}}=0.18\,r_0$, $w_{\mathrm{iC}}=0.018\,r_0$ and the outer with $r_{\mathrm{oC}}=7\,r_0$, $w_{\mathrm{oC}}=0.7\,r_0$.

When the simulation was initialised, we immediately released the embedded planet to migrate based on its interaction with the disc. For simplicity, we disregarded accretion onto the planet.

\subsection{Numerical setup}
\label{sec:setup}

Our numerical grid extends from $0.1\,r_0$ to $10\,r_0$ in the radial direction, with logarithmic cell spacing. In the azimuthal direction it extends from $0$ to $2\pi$, with evenly spaced cells. We resolved our grid at $32$\,cells~per~scale~height~(cps), which corresponds to $2108$\,cells in the radial and $2873$\,cells in the azimuthal direction at $h_0=0.07$. We carried out a resolution convergence study (see Appendix~\ref{app:resolution}) and found that simulations at this resolution yield results consistent with simulations performed at $64$\,cps, while lower resolutions can lead to divergent migration behaviour within the same overall migration regime.

For the outer boundary, we used the reflecting condition and implemented a damping zone at $r\geq8\,r_0$ as an exponential relaxation of the fluid parameters to the initial state. We find that the presence of an inner damping zone causes a numerical instability, producing a strongly excited planetary eccentricity (see Appendix~\ref{app:boundary}). To avoid this behaviour, we did not implement a damping zone at the inner boundary. To prevent any possible interference due to reflections or pileups in its absence, we set the inner boundary condition to outflow.

We ran our simulations for at least $900$\,orbits at $r_0$, corresponding to about $320\,\mathrm{kyr}$ for $r_0=50\,\mathrm{au}$. From then on, when the planet reached the region of the cutoff in the initial gas surface density profile at $r=9\,\mathrm{au}$, we terminated the simulation, as we could no longer rely on the physicality of the simulation this close to the boundary. Otherwise, we continued the simulations for up to $1500$\,orbits, equivalent to $530\,\mathrm{kyr}$.

\subsection{Model parameters}
\label{sec:parameters}
Based on this setup, we performed two sets of simulations, varying the disc's aspect ratio, $h_0$ (and therefore $T_0$), turbulent viscosity parameter, $\alpha$, and the planetary mass, $M_{\mathrm{p}}$. In the first set, we investigated the effect of the aspect ratio and viscosity on the migration behaviour of a planet with mass $M_{\mathrm{p}}=100\,\mathrm{M}_{\oplus}$ and the resulting formation of structures in the disc's gas surface density. To this end, we chose four values for $h_0$: $0.06$, $0.07$, $0.08$, and $0.09$, as well as three values for $\alpha$: $10^{-3}$, $10^{-4}$, and $10^{-5}$.

For the second set, we studied the influence of the planet mass in two discs with $h_0=0.07$ and $h_0=0.08$ at $\alpha=10^{-4}$. For this purpose, we chose a wide range of planet masses $M_{\mathrm{p}}$: $30\,\mathrm{M}_{\oplus}$, $50\,\mathrm{M}_{\oplus}$, $70\,\mathrm{M}_{\oplus}$, $100\,\mathrm{M}_{\oplus}$, $150\,\mathrm{M}_{\oplus}$, $240\,\mathrm{M}_{\oplus}$, $320\,\mathrm{M}_{\oplus}$, and $400\,\mathrm{M}_{\oplus}$.

\section{Results}
\label{sec:results}

\begin{figure}
  \centering
  \includegraphics[width=\columnwidth]{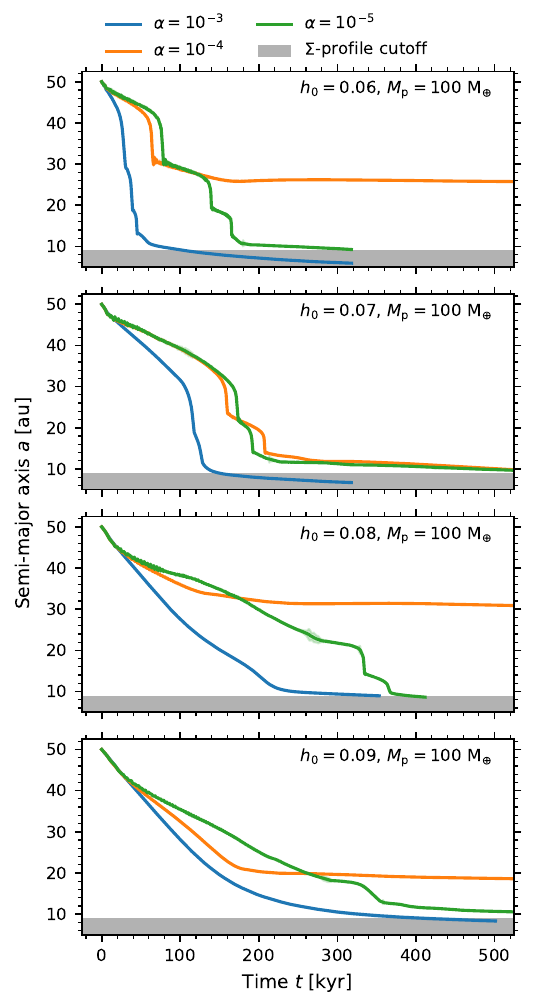}
  \caption{Migration tracks of the first set of simulations: Time evolution of the semi-major axis, $a$, of a $100\,\mathrm{M}_{\oplus}$ planet for three values of the turbulent viscosity parameter, $\alpha$, and four values of the disc aspect ratio, $h_0$. The presence and extent of a pale coloured region around a migration track (such as in the third panel, green track, around $260\,\mathrm{kyr}$) indicates whether and to what degree the planet's eccentricity is excited, as these regions span the extent between the planet's apsides. The grey band marks the cutoff in the initial gas surface density profile at $r=9\,\mathrm{au}$.}
  \label{fig:migration_tracks_set1}
\end{figure}

\subsection{Effect of aspect ratio and viscosity}
\label{sec:results_set1}
Figure~\ref{fig:migration_tracks_set1} shows the migration tracks of the first set of simulations, intended to gauge the influence of the disc's aspect ratio and viscosity on planet migration and structure formation. To check for instabilities similar to those triggered by the inner damping zone (see Appendix~\ref{app:boundary}), we looked for signs of a strongly excited planetary eccentricity. For this purpose, we included a pale coloured region around each migration track, spanning the planet's apsides. The presence and extent of this region directly indicate whether the eccentricity is excited, and by how much. Throughout this set of simulations, we find no signs of the instability. The planetary eccentricity is only occasionally mildly excited and tends to relax within a few kyr. This usually occurs when the planet transitions between different migration regimes or into and out of runaway migration episodes. Such minor and short-lived eccentricity perturbations are dynamically insignificant and do not alter the overall migration behaviour.

While migration is overall directed inward, we identified two distinct migration regimes, dependent on $h_0$, namely: intermittent type-III migration and smooth or vortex feedback. Both eventually transition into type-II migration, although the smooth- or vortex-feedback regime can, under suitable circumstances, also transition into the intermittent type-III regime.

\begin{figure*}
  \centering
  \includegraphics[width=\textwidth]{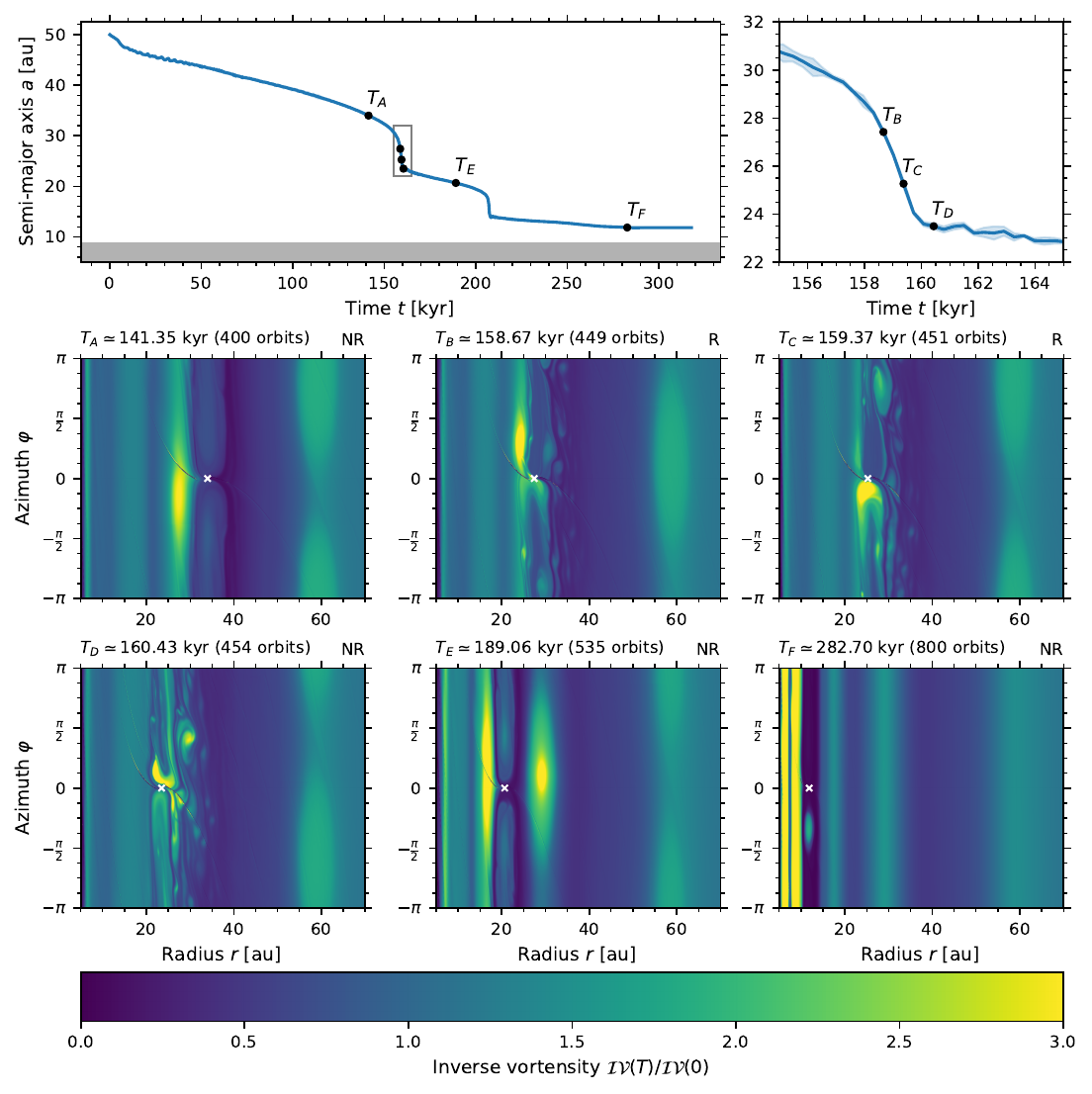}
  \caption{Intermittent migration of a $100\,\mathrm{M}_{\oplus}$ planet in a disc with $h_0=0.07$ and $\alpha=10^{-4}$. \textit{Top-left}: Time evolution of the semi-major axis, $a$, of the planet. The grey band marks the cutoff in the initial gas surface density profile at $r=9\,\mathrm{au}$, and the grey rectangle indicates a zoomed-in section (\textit{top-right}). \textit{Remaining panels}: Six screenshots of the inverse vortensity, $\mathcal{IV}=\Sigma/(\nabla\times v)_z$, between the inner boundary at $5\,\mathrm{au}$ and $70\,\mathrm{au}$, normalised to initial values, taken at times denoted by $T_A$ to $T_F$ in the upper panels. Above each $\mathcal{IV}$ panel, the time and corresponding number of orbital periods at the planet's initial location ($50\,\mathrm{au}$), as well as whether the current migration is runaway (R) or not (NR) is indicated. The planet's position is marked by the white cross.}
  \label{fig:diskevol_IV}
\end{figure*}

\subsubsection{Intermittent type-III regime}
For $h_0=0.06$ and $h_0=0.07$, we find that the planet alternates between episodes of 'slow' (although we note that, comparatively, these can be quite fast depending on the planet and disc parameters) and type-III runaway migration at all viscosities. This leads to a characteristic staircase shape in the migration tracks. The same behaviour was reported by \citetalias{2019MNRAS.484..728M} and \citetalias{2020MNRAS.493.5892W}, with the latter referring to it as 'intermittent migration'. \citetalias{2020MNRAS.493.5892W} also provided a detailed explanation of the underlying mechanisms through analysis of the time evolution of the inverse vortensity, $\mathcal{IV}$, around the planet in a disc with $\alpha=10^{-3}$. To better understand the migration process, the emerging structures and the effect of the disc viscosity on both, we reproduce their analysis for our $h_0=0.07$, $\alpha=10^{-4}$ model, and point out any differences. We take six snapshots of the $\mathcal{IV}$ at times $T_A$ to $T_F$ as displayed in Fig.~\ref{fig:diskevol_IV} and describe the migration process in the following:

\emph{$T_A$ -- slow migration ends:} Towards the end of the first slow migration episode, just before the onset of the first runaway, we find that the planet is located at $R_p=34\,\mathrm{au}$. It sits in an asymmetric partial gap, that is most strongly depleted at $R_\mathrm{dpl}\approx38\,\mathrm{au}$, just outside the corotation region towards the outer gap edge. As the planet migrates inward and viscous diffusion is slow to fill the gap, the partial gap is extended towards higher radii, featuring a gradient in $\mathcal{IV}$ ($R_\mathrm{ext}\sim40\text{--}55\,\mathrm{au}$). To either side of the gap region, $\mathcal{IV}$ maxima, which are ring-like in \citetalias{2020MNRAS.493.5892W}, are present in the form of azimuthally stretched, large-scale vortices. These are created by repeated shocks from the planet's wake \citep{2001ApJ...552..793G, 2010MNRAS.405.1473L}. However, such repeated shocks can only occur as long as the planet's migration timescale across the gap is longer than the synodic period of the gas at the gap edges \citepalias{2020MNRAS.493.5892W}. Consequently, the buildup at the outer edge is no longer sustained once the migration rate increases towards the end of the slow migration episode. This results in an observable asymmetry in both intensity and radial location relative to the planet for the two maxima, with the inner located at $R_\mathrm{imax}\approx28\,\mathrm{au}$ and the outer at $R_\mathrm{omax}\approx60\,\mathrm{au}$.

\emph{$T_B$ -- runaway migration:} As the planet continues to migrate inward, it begins to draw material from the inner gap edge $\mathcal{IV}$ maximum on outward horseshoe turns. As a result, $\delta m$ (see Eq.~\ref{eq:massdeficit}), and therefore the dynamical corotation torque and by extension the inward migration rate increases, until the planet can no longer capture the inflowing material in libration. With the high $\mathcal{IV}$ material now being flung past the planet on a single outward horseshoe turn, $\delta m$ increases drastically, and the planet enters into a type-III runaway migration episode (see Sect.~\ref{sec:theory}). During this process, a large amount of angular momentum is transferred from the planet to the disc material, rapidly moving the planet inward and the disc material outward. Due to the rapid inward migration, the planet is no longer able to sustain the inner $\mathcal{IV}$ maximum via repeated shocks, so that the runaway draws from a limited reservoir of high $\mathcal{IV}$ material. As the reservoir takes the shape of an asymmetric vortex structure rather than a symmetric ring structure as seen in \citetalias{2020MNRAS.493.5892W}, we do not observe a continuous stream of high $\mathcal{IV}$ material crossing the planet's orbit. Instead, pockets of material lower or higher in $\mathcal{IV}$ are drawn from the vortex, depending on the planet's azimuthal position relative to the centre of the vortex. However, due to the rapid nature of the runaway, this does not significantly affect the overall migration behaviour. Once a pocket of high $\mathcal{IV}$ material passes into the low $\mathcal{IV}$ region in the gap beyond the planet, the RWI, favoured by the low viscosity in our disc model, is triggered, leading to the formation of multiple small-scale vortices at the outer gap edge. This is another important difference to the $\alpha=10^{-3}$ model analysed in detail by \citetalias{2020MNRAS.493.5892W}, where the formation of vortices is suppressed due to the higher viscosity. It is, however, in line with their and \citetalias{2019MNRAS.484..728M} findings for discs at $\alpha<10^{-3}$.

\emph{$T_C$ -- runaway migration ends:} When the material with maximum $\mathcal{IV}$ gets flung outward past the planet, the planet reaches its maximum inward migration rate. However, from this point onward, the reservoir of high $\mathcal{IV}$ material is depleted. The planet now draws from low $\mathcal{IV}$ material, leading to a decrease in $\delta m$ and the migration rate, stopping the runaway.

\emph{$T_D$ -- horseshoe jostling:} Once the migration has slowed down sufficiently, some of the high $\mathcal{IV}$ material still left in the horseshoe region is trapped in libration. There it performs multiple inward and outward horseshoe turns in sequence, jostling the planet outward and inward. While also featuring the capture of high $\mathcal{IV}$ material in libration, this jostling is absent in the $\alpha=10^{-3}$ model of \citetalias{2020MNRAS.493.5892W}. We attribute this difference to the enhanced mixing of the horseshoe region in the latter. 

\emph{$T_E$ -- slow migration resumes:} Eventually, the libration region becomes sufficiently mixed and slow migration as seen at \emph{$T_A$} resumes. The planet once more opens an asymmetrical partial gap, building up a new $\mathcal{IV}$ maximum at the inner edge, $R_\mathrm{imax}\approx16\,\mathrm{au}$, due to repeated shocks from its wake. At the outer edge of the new gap, $R_\mathrm{omax}\approx29\,\mathrm{au}$, the vortices spawned during the runaway episode coalesce into a single large-scale vortex, making up the outer $\mathcal{IV}$ maximum. Beyond this vortex, the gap inhabited by the planet prior to the runaway and the original outer $\mathcal{IV}$ maximum persist at $R_\mathrm{gap1}\sim35\text{--}55\,\mathrm{au}$ and $R_\mathrm{max1}\approx58\,\mathrm{au}$, respectively. Henceforth, we refer to the structures left behind by the planet as 'remnants'.

\emph{$T_F$ -- stall after vortex smear-out:} Towards the end of the simulation, after another runaway episode, migration finally stalls. We suspect that the coorbital region is at this point too depleted to drive the planet into another runaway episode. The planet opens a deep gap at its location, $R_\mathrm{gap}\sim10\text{--}15\,\mathrm{au}$, with a dissipating high $\mathcal{IV}$ vortex trapped at the L5 Lagrange point and $\mathcal{IV}$ maxima at either gap edge. Due to the second runaway episode, an additional remnant gap and $\mathcal{IV}$ maximum are present at $R_\mathrm{gap2}\sim20\text{--}25\,\mathrm{au}$ and $R_\mathrm{max2}\approx29\,\mathrm{au}$, respectively. All remnant structures from previous runaways still persist at this point, though the large-scale vortices at the $\mathcal{IV}$ maxima have azimuthally smeared out into symmetric ring structures.

\begin{figure*}
  \centering
  \includegraphics[width=\textwidth]{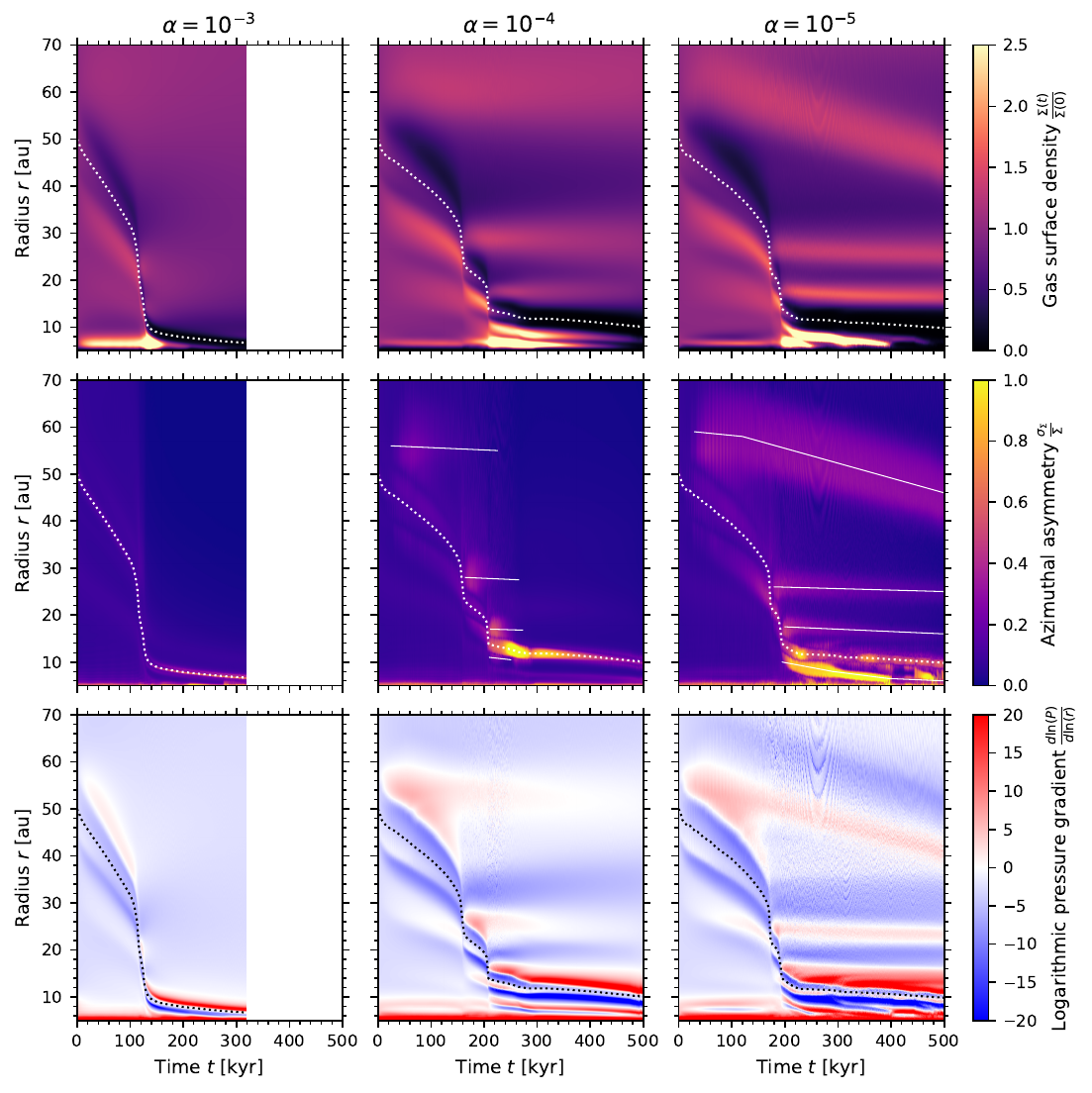}
  \caption{Azimuthally averaged time evolution of the gas for a $100\,\mathrm{M}_{\oplus}$ planet in a disc with $h_0=0.07$ at different $\alpha$: The dotted line represents the planet's migration track. \textit{Top row}: Azimuthally averaged gas surface density, $\Sigma$, normalised to initial values. \textit{Bottom row}: Logarithmic gas pressure gradient at the midplane. Stark white regions at the transition from red to blue (towards larger $r$) are indicative of local gas pressure maxima, likely candidates for dust traps. \textit{Middle row}: Azimuthal standard deviation of $\Sigma$ normalised to its current value at time, $t$. This illustrates the asymmetries in the gas surface density, which are indicative of vortices. Note: this quantity is only meaningful outside deep gaps. The relevant primary (inner and outer) and remnant vortices are highlighted by white lines tracing their approximate radial centres until their eventual smear-out into rings.}
  \label{fig:set1_h007_aziavg}
\end{figure*}

The formation of remnant structures due to intermittent migration enables a planet to form multiple dust-trapping pressure bumps at various distances to the star during its migration. Provided that these structures persist for a significant fraction of the disc's lifetime, this could account for the dust rings observed at large stellocentric radii in many protoplanetary discs. The survival of these structures is mainly determined by the disc viscosity, as higher viscosities work to smooth out pressure bumps. However, at lower viscosities, the RWI is favoured and formed vortices endure for longer timescales. As vortices are much less common than axisymmetric dust rings in ALMA observations, it becomes crucial to find a viscosity regime where vortices quickly smear out azimuthally, while the resulting ring-like structures persist for long timescales afterwards.

Figure~\ref{fig:set1_h007_aziavg} gives an overview of the evolution of the emerging structures in the gas surface density for $h_0=0.07$ at the three viscosities, tracing the azimuthally averaged gas surface density, $\Sigma$ (top row), and the resulting logarithmic gas pressure gradient at the midplane, $d\operatorname{ln}(P)/d\operatorname{ln}(r)$ (bottom row). Pressure maxima, which act as dust traps, show in the plot as stark white regions at the transition from red to blue (towards larger $r$). As we do not explicitly evolve dust in our models, we infer the presence and survival of dust rings from the location, strength, and longevity of these pressure maxima. To indicate the presence of vortices, which make up most if not all asymmetries in $\Sigma$ in our simulations, Fig.~\ref{fig:set1_h007_aziavg} also includes the standard deviation of $\Sigma$ in the azimuthal direction normalised to its current value at time, $t$ (middle row). We dub this quantity the azimuthal asymmetry of $\Sigma$. The azimuthal asymmetry allows us to identify whether overdensities are vortex-like (strong asymmetry) or ring-like (weak asymmetry). It should be noted, however, that this quantity is only meaningful outside deep gaps. The large values inside the planetary gaps therefore have no particular meaning. While the azimuthal asymmetry provides a good indicator for the presence of vortices, we find that it can lack precision when determining smear-out timescales. To obtain more precise smear-out timescales, we cross-reference with the time evolution of the $\mathcal{IV}$ in $r$ and $\varphi$.

The first column in Fig.~\ref{fig:set1_h007_aziavg} shows the high-viscosity model at $\alpha=10^{-3}$. We find that the planet undergoes two runaway migration episodes, the first one after $T_{\mathrm{rm}1}\approx110\,\mathrm{kyr}$ at $R_{\mathrm{rm}1}\approx27\,\mathrm{au}$, the second after $T_{\mathrm{rm}2}\approx120\,\mathrm{kyr}$ at $R_{\mathrm{rm}2}\approx16\,\mathrm{au}$. While both leave behind a remnant gap with an overdensity at the outer edge, these structures already strongly diffuse within a few kyr after the planet migrates away, and are completely smoothed over within $\lesssim100\,\mathrm{kyr}$. This is especially true for structures left by the second runaway, which only had a very brief preceding slow migration episode to build up. The corresponding pressure maxima are located at $R_{\mathrm{r2Pmax}}\approx38\,\mathrm{au}$ and $R_{\mathrm{r1Pmax}}\approx20\,\mathrm{au}$, and likewise vanish within a few kyr. After the second runaway, the planet transitions to type-II migration and stalls in a deep gap at $R_\mathrm{stall}\approx7\,\mathrm{au}$. The only pressure maxima that persist are located at the inner ($R_{\mathrm{iPmax}}\approx6\,\mathrm{au}$) and outer ($R_{\mathrm{oPmax}}\approx9\,\mathrm{au}$) edge of this gap, where the tidal forces of the planet continue to act against viscous diffusion. Therefore, while the azimuthal asymmetry shows that vortices are mostly absent throughout the simulation, which is in agreement with the findings of \citetalias{2020MNRAS.493.5892W}, we find that the remnant lifetimes in this viscosity regime are simply too short for the observed dust rings to be the result of intermittent migration remnants.

In the second column, we present the model with moderately low viscosity, $\alpha=10^{-4}$, which we already analysed in detail in Fig.~\ref{fig:diskevol_IV}. Again, the planet undergoes two runaway migration episodes at similar radii, the first one after $T_{\mathrm{rm1}}\approx156\,\mathrm{kyr}$ at $R_{\mathrm{rm1}}\approx28\,\mathrm{au}$, the second after $T_{\mathrm{rm2}}\approx206\,\mathrm{kyr}$ at $R_{\mathrm{rm2}}\approx17\,\mathrm{au}$. At this viscosity, the remnant gap and outer edge overdensity left by each runaway persist, though slightly diffused towards the end, for the entire duration of the simulation (here plotted: $500\,\mathrm{kyr}$ out of $530\,\mathrm{kyr}$), even after the planet has migrated away from them. We find the corresponding pressure maxima at $R_{\mathrm{r2Pmax}}\approx54\,\mathrm{au}$ and $R_{\mathrm{r1Pmax}}\approx26\,\mathrm{au}$. While these are gradually smoothed out, they are likewise still present at the end of our simulation, giving these remnants a lifetime of at least $300\text{--}500\,\mathrm{kyr}$. Once again, after the second runaway, the planet transitions to type-II migration and stalls in a deep gap, this time at $R_\mathrm{stall}\approx10\,\mathrm{au}$, generating strong, persistent pressure maxima at the inner ($R_{\mathrm{iPmax}}\approx8\,\mathrm{au}$) and outer ($R_{\mathrm{oPmax}}\approx16\,\mathrm{au}$) gap edge, respectively. The azimuthal asymmetry shows the initial presence of vortices at the gap edges, where the overdensities and pressure maxima are located. We find that these vortices smear out prior to the end of the simulation, on a timescale that is larger farther out in the disc: the outermost vortex left by the first runaway smears out about $\Delta T_\mathrm{r1Vso}\approx200\,\mathrm{kyr}$ after its formation, the vortex left by the second runaway after $\Delta T_\mathrm{r2Vso}\approx100\,\mathrm{kyr}$, the vortex at the outer gap edge of the stalled planet after $\Delta T_\mathrm{oVso}\approx60\,\mathrm{kyr}$, and the vortex at the inner gap edge after $\Delta T_\mathrm{iVso}\approx40\,\mathrm{kyr}$. We conclude that the viscosity in this regime is low enough to allow for the long-term survival of the underlying pressure maxima of remnant structures, while still being sufficient to facilitate the smear-out of their constituent vortices into axisymmetric ring structures on a several times shorter timescale.

The low-viscosity model at $\alpha=10^{-5}$, as shown in the third column, shows once again two runaway migration episodes at similar radii to the other two models: the first after $T_{\mathrm{rm1}}\approx170\,\mathrm{kyr}$ at $R_{\mathrm{rm1}}\approx28\,\mathrm{au}$, the second after $T_{\mathrm{rm2}}\approx189\,\mathrm{kyr}$ at $R_{\mathrm{rm2}}\approx18\,\mathrm{au}$. The corresponding gaps and overdensities form, which, due to the much lower viscosity compared to the previous models, show little to no signs of diffusion, even at the end of our simulations. The underlying pressure maxima are located at $R_{\mathrm{r2Pmax}}\approx60\text{--}44\,\mathrm{au}$ and $R_{\mathrm{r1Pmax}}\approx25\,\mathrm{au}$, respectively. We note that the outer pressure maximum starts out farther outward and moves inward, presumably due to migration of the massive outer vortex at its location \citep{2010ApJ...725..146P}. The stall occurs at $R_\mathrm{stall}\approx10\,\mathrm{au}$, generating strong, persistent pressure maxima at $R_{\mathrm{iPmax}}\approx7\,\mathrm{au}$ and $R_{\mathrm{oPmax}}\approx16\,\mathrm{au}$. While the lower viscosity clearly increases the lifetime expectations for structures beyond the duration of our simulations, the azimuthal asymmetry reveals that the smear-out timescales similarly increase, with only the inner vortices starting to smear out towards the end of the simulation.

In summary: the stall after the occurrence of intermittent type-III migration typically occurs at radii too small to account for any but the innermost rings of ALMA observations \citep{2018ApJ...869L..42H} and close enough to our simulation boundary to warrant caution. Remnants left by rapid migration episodes can account for multiple rings farther out in the disc, but only survive for longer times at low viscosities $\alpha\leq10^{-4}$. Lower viscosity leads to longer lifetimes but also increases the time these structures spend as vortices. A viscosity of $\alpha=10^{-4}$ is best suited to showcase the vortex smear-out within the duration of our simulations.

\subsubsection{Smooth- or vortex-feedback regime}
Returning to Fig.~\ref{fig:migration_tracks_set1}, we find, in the case of $h_0=0.08$ and $h_0=0.09$, that after a brief initial period of type-I migration (a few kyr), the planet opens a partial gap in the gas surface density. As the planet migrates farther inward, the gap develops an asymmetry, opening deeper at the outer side of the planet's orbit than at the inner. This leads to a reduction of the negative torque acting on the planet due to the outer Lindblad resonances. As a result, the migration rate decreases, until the planet eventually transitions to type-II migration and stalls in a deep gap.

As gap opening efficiency is increased for lower $h_0$ (lower $M_\mathrm{th}$), we expect the stall to occur earlier and at larger radii for $h_0=0.08$ compared to $h_0=0.09$. This behaviour is clearly visible for $\alpha=10^{-4}$. It is less obvious for $\alpha=10^{-3}$, where the gap opening efficiency is much more affected by the higher viscosity working to refill the gap, leading to a much smaller decrease in migration rate overall. For $\alpha=10^{-5}$ this relation is entirely broken, as the lower viscosity is no longer able to effectively suppress vortices.

Despite enhanced gap-opening due to the lower viscosity, inward migration is eventually assisted by small-scale vortices formed through the RWI at the outer gap edge. As a result, the migration rate periodically stabilises or even increases. Furthermore, the planet is intensely jostled by vortices during the early stages of migration, $t \lesssim 200\,\mathrm{kyr}$, as seen in the erratic migration track. Lastly, once migration begins to stall as the planet deepens the partial gap at its location, a large vortex forms at the outer gap edge, driving renewed inward migration until the planet eventually transitions into type-II migration and stalls in a deep gap. As seen for $h_0=0.08$ at $\alpha=10^{-5}$, this process can even push the planet into the intermittent type-III regime. 

It should be noted, that while the RWI also spawns small-scale vortices at the outer gap edge at $\alpha=10^{-4}$, they are much weaker and shorter lived compared to $\alpha=10^{-5}$. We find that these do not significantly affect the migration behaviour beyond small oscillations in the migration track at early times and slight alterations to the migration rate due to weak vortex-assisted migration.

An overview of the azimuthally averaged time evolution of the gas for the $h_0=0.08$ simulations in the manner of Fig.~\ref{fig:set1_h007_aziavg} can be found in Fig.~\ref{fig:set1_h008_aziavg}.

Overall, the migration behaviour for $h_0=0.08$ and $h_0=0.09$ is consistent at high viscosity with the 'smooth feedback' regime and at low viscosity with the 'feedback with vortices' regime outlined by \citetalias{2019MNRAS.484..728M}. In the latter case, episodes of both vortex-assisted migration by short-lived, small-scale vortices \citepalias{2019MNRAS.484..728M} and vortex-driven migration by a single long-lived, large-scale vortex \citep{2022A&A...658A..32L} can be observed. In all cases, inward migration eventually stalls, allowing the planet to continuously support a dust-trapping pressure bump at similar radii for a significant fraction of the disc's lifetime. However, at $\alpha=10^{-3}$ and $\alpha=10^{-5}$, this stall occurs too close to the star ($r \lesssim 10\,\mathrm{au}$) to explain the dust rings observed farther out in the disc and close enough to the boundary of our simulations to warrant caution. On the other hand, the model at $\alpha=10^{-4}$ shows that stalls at higher stellocentric radii are possible, with the planet stalling at about $R_\mathrm{stall}\approx31\,\mathrm{au}$ for $h_0=0.08$ and at about $R_\mathrm{stall}\approx19\,\mathrm{au}$ for $h_0=0.09$.

\subsection{Effect of the planet mass}
\label{sec:results_set2}
The disc parameters for the second set of simulations, aimed at investigating the influence of the planet mass on migration behaviour and structure formation, were narrowed down based on the results of the first set. We chose two aspect ratios, $h_0=0.07$ and $h_0=0.08$, to cover both of the migration regimes which we found in our first set. For the viscosity, we selected $\alpha=10^{-4}$, as it allows for the formation of long-lived remnant structures in the intermittent regime, while keeping the vortex smear-out timescale within the simulation duration.

\begin{figure}
  \centering
  \includegraphics[width=\columnwidth]{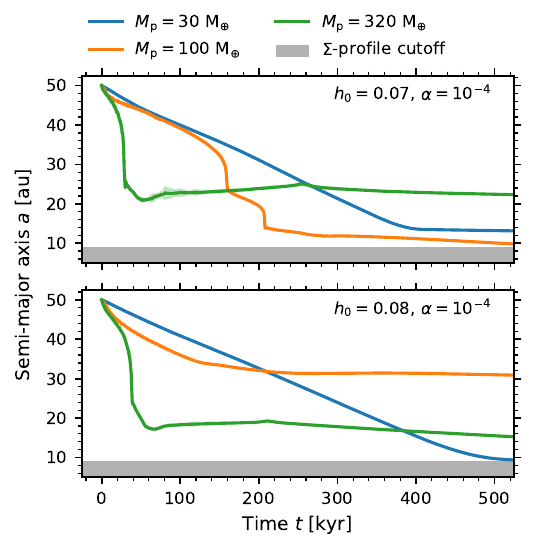}
  \caption{Selection of migration tracks from the second set of simulations: Time evolution of the semi-major axis, $a$, of planets with masses $30\,\mathrm{M}_{\oplus}$, $100\,\mathrm{M}_{\oplus}$, and $320\,\mathrm{M}_{\oplus}$ for two values of the disc aspect ratio, $h_0$, at $\alpha=10^{-4}$. The presence and extent of a pale coloured region around a migration track indicates whether and to what degree the planet's eccentricity is excited, as these regions span the extent between the planet's apsides. The grey band marks the cutoff in the initial gas surface density profile at $r=9\,\mathrm{au}$.}
  \label{fig:migration_tracks_set2_selection}
\end{figure}

For better readability, only a selection ($M_{\mathrm{p}}=30\,\mathrm{M}_{\oplus}$, $100\,\mathrm{M}_{\oplus}$, $320\,\mathrm{M}_{\oplus}$) of the migration tracks from this second set of simulations is presented in Fig.~\ref{fig:migration_tracks_set2_selection}. At $h_0=0.07$, these are representative of the three migration regimes emerging at different planet masses. A complete overview of the migration tracks for all planet masses (see Sect.~\ref{sec:parameters}), most only spanning a simulation duration of $320\,\mathrm{kyr}$, is provided in Fig.~\ref{fig:migration_tracks_set2}.

As with the first set of simulations, we included a pale coloured region spanning the planet's apsides around each migration track, which allows us to diagnose any excitation of the planetary eccentricity. The second set likewise shows no evidence of the strongly excited eccentricities associated with the numerical instability seen in simulations with an inner damping zone. Instead, we again find only occasional, minor, and short-lived eccentricity perturbations that are dynamically insignificant and do not affect the overall migration behaviour. As in the first set, these brief excitations typically occur during transitions between migration regimes and at the beginning or end of runaway episodes.

\subsubsection{Mapping migration regimes at different planet masses}
We find that for $M_{\mathrm{p}}\lesssim 70\,\mathrm{M}_{\oplus}$ (represented in Fig.~\ref{fig:migration_tracks_set2_selection} by the $30\,\mathrm{M}_{\oplus}$ planet), the planet migrates inward in the feedback regime for both aspect ratios. Eventually, the planet transitions to type-II migration and stalls in a deep gap at $R_\mathrm{stall}\sim10\text{--}40\,\mathrm{au}$. In this regime, we find that stalls occur earlier and farther outward for higher $M_{\mathrm{p}}$ and later and farther inward for higher $h_0$. There are instances of jostling by short-lived, small-scale vortices spawned by the RWI at the outer edge of the partial gap over the first $100\,\mathrm{kyr}$, but we observe no episodes of vortex-assisted migration that significantly impact the overall migration behaviour.

As already discussed, the migration behaviour diverges for the two disc aspect ratios at a planet mass around $100\,\mathrm{M}_{\oplus}$: for $h_0=0.07$, a planet with $M_{\mathrm{p}}=100\,\mathrm{M}_{\oplus}$ migrates in the intermittent type-III regime, while for $h_0=0.08$, it migrates in the smooth- or vortex-feedback regime. For $M_{\mathrm{p}}\gtrsim150\,\mathrm{M}_{\oplus}$, migration then wholly shifts into the intermittent type-III regime across both values of $h_0$. In this set, planets undergoing intermittent migration consistently feature two runaway episodes, with the final stall occurring between $R_\mathrm{stall}\sim10\text{--}20\,\mathrm{au}$. Again, we find that the transition to type-II migration occurs earlier and farther outward for higher $M_{\mathrm{p}}$ compared to other planets migrating in the same regime. The same is true for the onset of the runaway episodes, despite the duration of the intermediate slow migration episodes increasing with higher $M_{\mathrm{p}}$. While we have only limited data to identify a relation for $h_0$ in the intermittent regime, we find that higher $h_0$ leads to a later onset of the first runaway, shorter intermediate slow migration episodes, and a later and farther inward occurring transition to type-II migration. We note that compared to the feedback regime at lower masses, where the planet's migration stalls on transition to type-II migration, planets in the intermittent regime still noticeably trend inward in a near-stall type-II manner.

Eventually, for even higher $M_{\mathrm{p}}$, the migration behaviour of planets in the two disc models diverges again: at around $M_{\mathrm{p}}=240\,\mathrm{M}_{\oplus}$, the planet in the disc with $h_0=0.08$ continues to migrate in the intermittent type-III regime, while the planet in the disc with $h_0=0.07$ displays a similar, yet slightly distinct migration behaviour. At $M_{\mathrm{p}}\gtrsim320\,\mathrm{M}_{\oplus}$, here represented by the $320\,\mathrm{M}_{\oplus}$ planet, we observe this new migration regime for both aspect ratios. It is characterised by an initial period of slow migration leading directly into a single type-III runaway episode, which is followed by another slow migration episode that transitions into a type-II-like outward migration \citep{2022MNRAS.514.5478S}. Eventually, this outward migration reverses, and the planet trends inward in a near-stall type-II manner, at a distance to the central star of about $10\,\mathrm{au}$ to $30\,\mathrm{au}$. Given its similarity to the intermittent migration regime, bar the different behaviour in near-stall, we consider it a high-mass variant of the same, rather than an entirely new regime. While our data is again limited, the $M_{\mathrm{p}}$ and $h_0$ relations for the migration behaviour do indeed appear to be similar to those established for the intermittent type-III regime, with the noticeable exception of the $M_{\mathrm{p}}=400\,\mathrm{M}_{\oplus}$ simulation, which for $h_0=0.08$ even lacks the runaway episode during the inward migration prior to the near-stall.

\subsubsection{Transition to the intermittent type-III migration}

\begin{figure}
  \centering
  \includegraphics[width=\columnwidth]{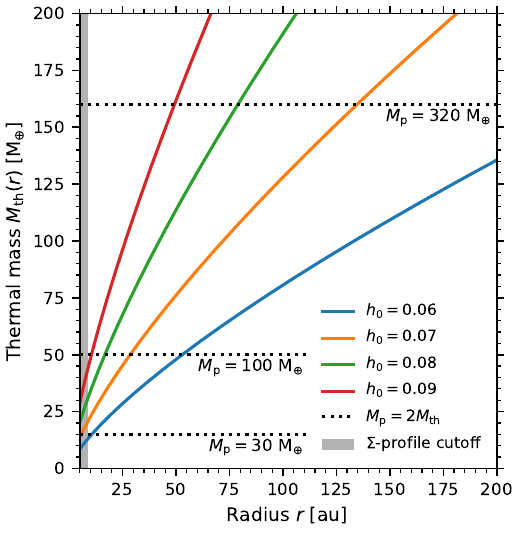}
  \caption{Transition thresholds for intermittent type-III migration: Thermal mass $M_\mathrm{th}(r)$ (Eq.~\ref{eq:thermalmass}) profiles of our four disc models ($h_0$, as $M_\mathrm{th}$ is independent of $\alpha$). The dotted lines mark the postulated approximate thresholds for transition to intermittent type-III migration, $M_\mathrm{p}\approx2M_\mathrm{th}$, for the selection of planet masses featured in Sec.~\ref{sec:results_set2}. The grey band marks the cutoff in the initial gas surface density profile at $r=9\,\mathrm{au}$.}
  \label{fig:thermal_mass}
\end{figure}

We reason that the shift in migration behaviour from smooth or vortex feedback to intermittent type-III with increasing planet mass and lower aspect ratio should imply the existence of a critical planet mass similar to the feedback (see Eq.~\ref{eq:feedbackmass}) or thermal mass (see Eq.~\ref{eq:thermalmass}) criterion. Indeed, we find that in our simulations the onset of the first runaway migration typically occurs at around $M_{\mathrm{p}}\gtrsim 2 M_{\mathrm{th}}$, while planets that do not experience runaway migration stall prior to reaching disc regions where this condition is met (remember that for our flared disc $M_{\mathrm{th}}\propto r^{3/4}$, see Eq.~\ref{eq:thermalmass} and Eq.~\ref{eq:aspectratio_temperatureprofile}). This is illustrated in Fig.~\ref{fig:thermal_mass}, which shows the thermal mass profile $M_{\mathrm{th}}(r)$ for all disc models and the corresponding critical mass thresholds for the selection of planet masses featured in this section. However, without a means to identify under which parameters planets stall prior to meeting this criterion, it is not a particularly useful predictor for the migration behaviour.

\begin{figure}
  \centering
  \includegraphics[width=\columnwidth]{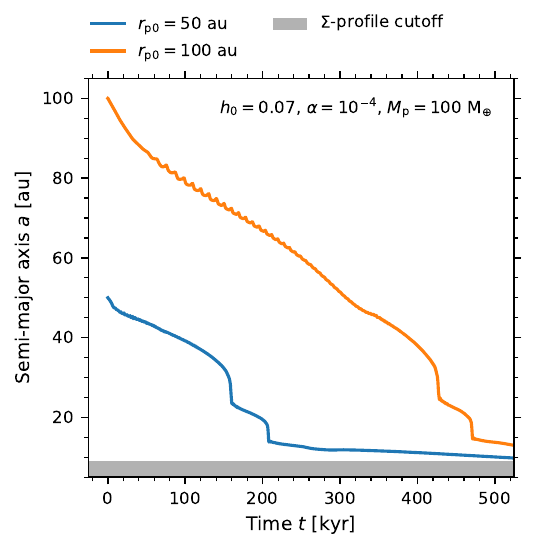}
  \caption{Same as Fig.~\ref{fig:migration_tracks_set2_selection} but for a $100\,\mathrm{M}_{\oplus}$ planet in a disc with $h_0=0.07$, $\alpha=10^{-4}$ and initial planet positions, $r_\mathrm{p0}$, at $50\,\mathrm{au}$ and $100\,\mathrm{au}$.}
  \label{fig:migration_tracks_onset100}
\end{figure}

Even so, to confirm that it is indeed the migration into a disc region where $M_{\mathrm{p}}\gtrsim 2 M_{\mathrm{th}}$ that triggers the onset of runaway migration, we performed an additional simulation of a $100\,\mathrm{M}_{\oplus}$ planet in a disc with $h_0=0.07$ and $\alpha=10^{-4}$, but this time embedding the planet at an initial distance of $r_\mathrm{p0}=100\,\mathrm{au}$, while keeping $r_0=50\,\mathrm{au}$. Figure~\ref{fig:migration_tracks_onset100} compares the resulting migration track to the original simulation. As expected, the planet that was embedded at $100\,\mathrm{au}$ initially migrates inward in the vortex-feedback regime, featuring an elongated episode of intensive jostling by vortices, until it switches to the intermittent type-III regime at a similar stellocentric radius to the original simulation.

\begin{figure}
  \centering
  \includegraphics[width=\columnwidth]{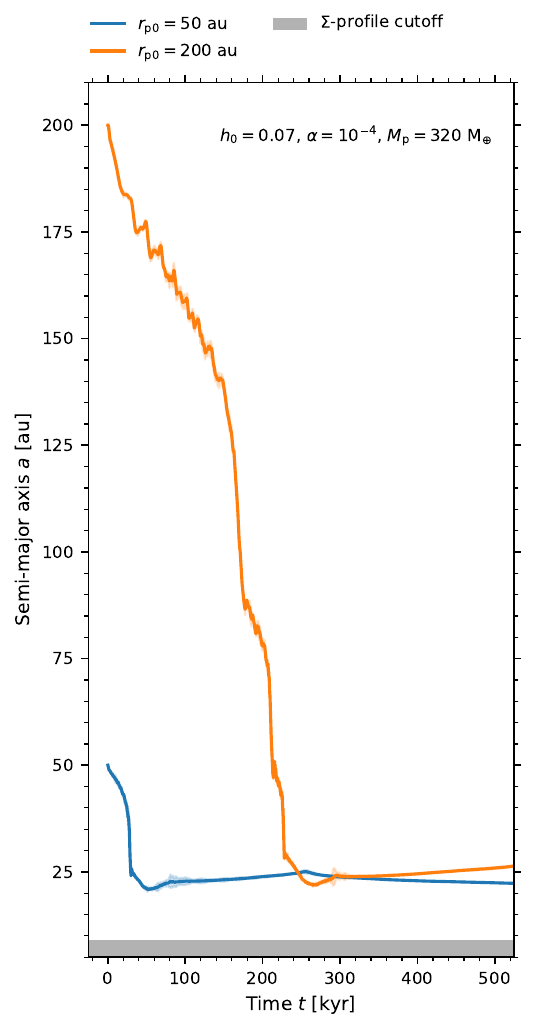}
  \caption{Same as Fig.~\ref{fig:migration_tracks_set2_selection} but for a $320\,\mathrm{M}_{\oplus}$ planet in a disc with $h_0=0.07$, $\alpha=10^{-4}$ and initial planet positions, $r_\mathrm{p0}$, at $50\,\mathrm{au}$ and $200\,\mathrm{au}$.}
  \label{fig:migration_tracks_onset320}
\end{figure}

\subsubsection{Massive planets initialised at large stellocentric radii}
If the critical planet mass for the onset of runaway migration is indeed around $M_{\mathrm{p}}\gtrsim 2 M_{\mathrm{th}}$, this raises an interesting point regarding the high-mass variant of the intermittent type-III regime: these planets already fulfil this criterion at much higher stellocentric radii than $r_0=50\,\mathrm{au}$. Consequently, we speculate that some of their variant features may not only be a result of the high planet mass, but also of the fact that they are already in runaway-favourable disc regions at the start of the simulation.

To test this hypothesis, we performed another simulation of a $320\,\mathrm{M}_{\oplus}$ planet in a disc with $h_0=0.07$ and $\alpha=10^{-4}$, but this time embedding the planet at an initial distance of $r_\mathrm{p0}=200\,\mathrm{au}$, while keeping $r_0=50\,\mathrm{au}$. Figure~\ref{fig:migration_tracks_onset320} compares the resulting migration track to the original simulation. As expected, the planet that was embedded at $200\,\mathrm{au}$ already enters into a runaway at $R_\mathrm{rm1}\approx135\,\mathrm{au}$, where $M_{\mathrm{p}}\approx 2 M_{\mathrm{th}}$. This occurs after an initial episode of slow migration featuring intensive jostling by RWI-spawned vortices. Following the first runaway episode, we observe two further phases of a slow migration episode followed by a runaway episode at $R_\mathrm{rm2}\approx73\,\mathrm{au}$ and $R_\mathrm{rm3}\approx42\,\mathrm{au}$, respectively. The last runaway is similar to the singular runaway episode in the original simulation. The subsequent transition into the near-stall featuring outwards type-II-like migration also occurs in a similar way to the original simulation.

We conclude that while this test once more confirms the mass criterion for the onset of runaway migration and that high-mass planets do indeed fall into the intermittent type-III regime, their near-stall behaviour presumably remains rooted in their high mass, setting them apart from lower mass planets in the same regime. As the runaways leave pressure maxima that persist for the entire duration of our simulations at $R_\mathrm{r1Pmax}\approx150\,\mathrm{au}$, $R_\mathrm{r2Pmax}\approx90\,\mathrm{au}$, and $R_\mathrm{r3Pmax}\approx55\,\mathrm{au}$, this simulation also illustrates that intermittent migration of massive planets which start farther out in the disc can generate rings out to large stellocentric distances.

\begin{figure*}
  \centering
  \includegraphics[width=\textwidth]{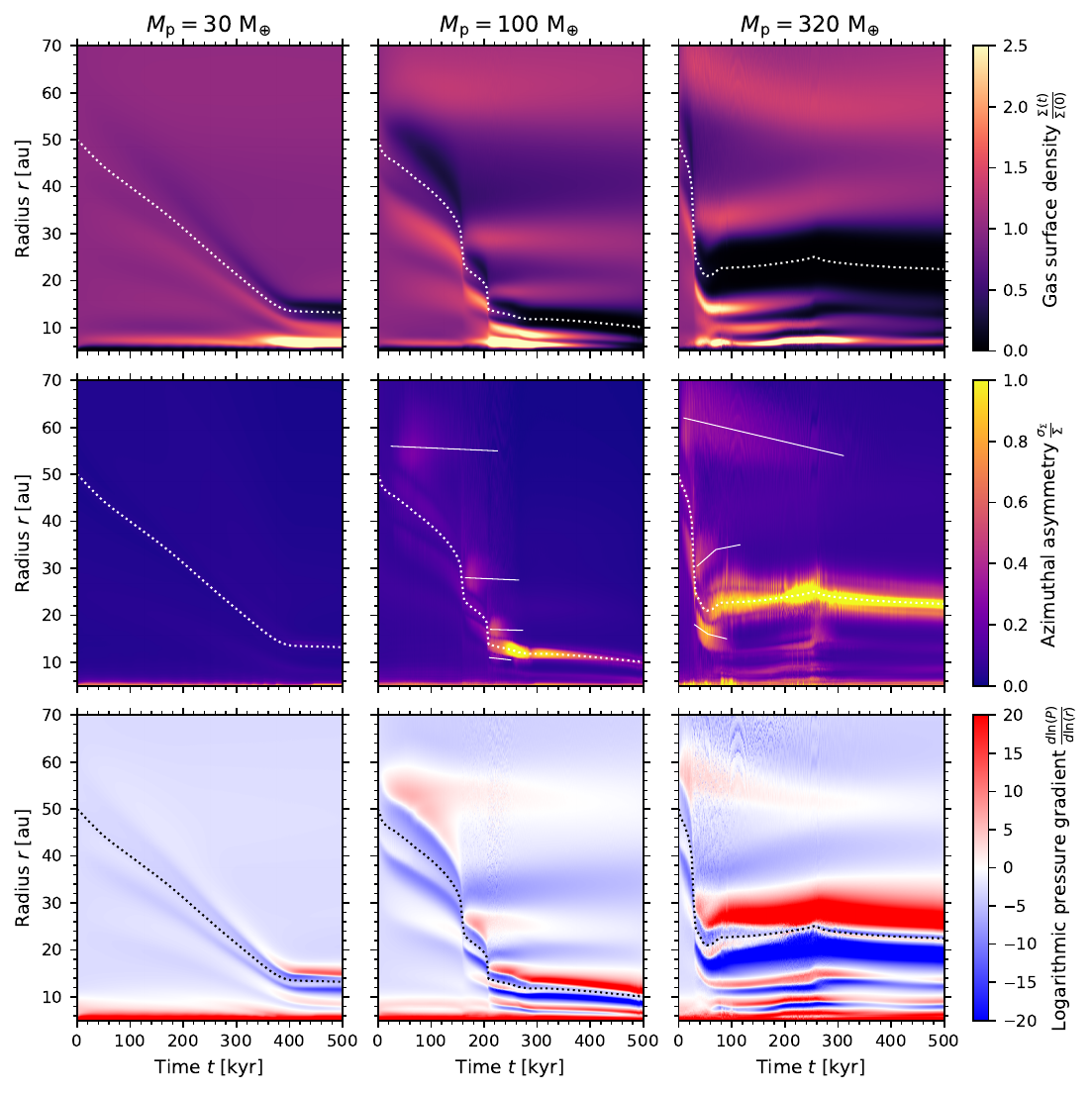}
  \caption{Same as Fig.~\ref{fig:set1_h007_aziavg} but for various planet masses in a disc with $h_0=0.07$ at $\alpha=10^{-4}$. For comparison, the middle columns in this figure and Fig.~\ref{fig:set1_h007_aziavg} depict the same simulation.}
  \label{fig:set2_h007_aziavg}
\end{figure*}

\subsubsection{Azimuthally averaged disc structure evolution}
For each migration regime, the emerging disc structures can be gleaned from the overview of the azimuthally averaged time evolution of the gas, plotted in Fig.~\ref{fig:set2_h007_aziavg} for $h_0=0.07$ at $\alpha=10^{-4}$ and planet masses of $30\,\mathrm{M}_{\oplus}$, $100\,\mathrm{M}_{\oplus}$, and $320\,\mathrm{M}_{\oplus}$, which are representative of the three migration regimes. The middle column in this figure depicts the same simulation as the middle column in Fig.~\ref{fig:set1_h007_aziavg}, which was already described in detail in Sect.~\ref{sec:results_set1}.

The left column shows the $30\,\mathrm{M}_{\oplus}$ planet, migrating in the feedback regime with vortices mostly absent. During its steady inward migration, the planet opens a partial gap, featuring weak pressure maxima at either gap edge. These structures migrate inward along with the planet. As the planet transitions to type-II migration and stalls in a deep gap at $R_\mathrm{stall}\approx13\,\mathrm{au}$, the local pressure maxima located at the inner ($R_\mathrm{iPmax}\approx10\,\mathrm{au}$) and outer ($R_\mathrm{oPmax}\approx17\,\mathrm{au}$) gap edge become more pronounced. Being directly supported by the stalled planet, they may persist for a significant fraction of the disc's lifetime, even beyond the duration of our simulations. Due to the absence of type-III runaway migration episodes, no remnant structures are formed at this mass.

Featured in the right column is the $320\,\mathrm{M}_{\oplus}$ planet, migrating in the high-mass variant of the intermittent type-III regime. Despite the initial slow migration quickly transitioning into a runaway, a remnant partial gap remains around the planet's initial location, with a pressure maximum at its outer edge, $R_\mathrm{r1Pmax}\approx52\,\mathrm{au}$. While showing clear signs of diffusion, the remnant persists for the entire duration of the simulation. The gap in which the planet finally comes to a near-stall at $R_\mathrm{stall}\approx22\,\mathrm{au}$ is deep and wide (at least $12\,\mathrm{au}$, not including the diffuse edges), with strong, persistent pressure maxima at its inner ($R_\mathrm{iPmax}\approx15\,\mathrm{au}$) and outer ($R_\mathrm{oPmax}\approx35\,\mathrm{au}$) edge. Interestingly, during the near-stall, we observe the formation of two smaller gaps at $R_\mathrm{gap1}\approx13\,\mathrm{au}$ and $R_\mathrm{gap2}\approx9\,\mathrm{au}$ inside the planet's orbit. Similar secondary gaps have been observed by \citet{2017ApJ...843..127D}, \citet{2017ApJ...850..201B}, and \citet{2018ApJ...869L..47Z}. The two gaps feature a pressure maximum at their shared edge. The outer of the two also shares a pressure maximum with the inner edge of the planet's primary gap, and eventually merges with the latter as the planet continues its slow inward migration during the near-stall. Any structures at the inner edge of the innermost gap cannot be reliably identified, as they are masked by the pressure maxima invariably produced at the inner boundary of our simulations.

This boundary feature is present throughout all of our simulations, and is the reason why the emergence of similar secondary structures inside the planet's orbit has so far been disregarded in our analysis: while the simulations at $\alpha=10^{-4}$ and $\alpha=10^{-5}$ in Fig.~\ref{fig:set1_h007_aziavg} both show the formation of a small secondary gap with corresponding pressure bumps at $R_\mathrm{gap1}\approx13\,\mathrm{au}$, they are located too close to the inner boundary to be reliably identified. We attribute the emergence of these secondary structures in low-viscosity discs to shocks from secondary and tertiary Lindblad spirals launched by the stalled planet \citep{2018ApJ...859..118B, 2018ApJ...859..119B}. Being directly supported by the planet, we expect these secondary structures to endure well beyond the duration of our simulations along with the primary gap, provided they do not merge with the latter as seen here.

We find that these secondary structures start out as vortices stretched out across almost the entire azimuth, which smear out within a few kyr. They briefly revert to vortices when the outward type-II-like migration reverses, before smearing out again. Note that this behaviour is difficult to distinguish in the azimuthal asymmetry plot due to the presence of meaningless asymmetry bands at the location of the secondary gaps. The vortices at the edges of the planet's gap, on the other hand, smear out after $\Delta T_\mathrm{iVso}\approx60\,\mathrm{kyr}$ and $\Delta T_\mathrm{oVso}\approx80\,\mathrm{kyr}$, while the outer vortex smears out after $\Delta T_\mathrm{r1Vso}\approx300\,\mathrm{kyr}$.

A further overview of the azimuthally averaged time evolution of the gas for the $h_0=0.08$ simulations can be found in Fig.~\ref{fig:set2_h008_aziavg}.

\section{Discussion}
\label{sec:discussion}

\subsection{Migrating planets as the origin of dust rings}
\label{sec:implications}
This study investigates whether migrating planets can account for long-lived, multi-ringed structures extending out to $r\sim150\,\mathrm{au}$, such as those seen with ALMA in numerous protoplanetary discs \citep{2018ApJ...869L..42H}. In our simulations, we identified three concurrent mechanisms capable of producing such rings. Here, we discuss their dynamics, significance, and implications.

\subsubsection{Rings due to a stalled planet}
Throughout all our simulations, we find that the planet eventually stalls as it opens a deep gap. Pressure maxima are located on both edges of this gap. Their potential to trap dust in rings is well documented in the literature \citep{2018ApJ...869L..47Z, 2021A&A...647A.174R, 2023ApJ...945L..37G, 2023MNRAS.524.3930Z}. For a deep gap to open, the planet's tidal forces must be strong enough to overcome both the disc's pressure and viscous forces \citep{2002ApJ...569..997R}. We see this reflected in our simulations, where the planets stall farther outward for higher planet mass (stronger tidal forces), lower aspect ratio (weaker pressure forces), and lower viscosity (lower viscous forces). However, we do find that there are two limiting factors to how far outward a planet can stall.

The first is the transition from a regime of smooth or vortex feedback \citepalias{2019MNRAS.484..728M} to one of intermittent type-III migration \citepalias{2019MNRAS.484..728M, 2020MNRAS.493.5892W}, as seen in Fig.~\ref{fig:migration_tracks_onset100}. While lower mass planets stall prior to this transition, higher mass planets experience one or more rapid migration episodes before they finally stall much farther inward than might be expected from their mass. We find that this transition occurs at a planet mass between $70\,\mathrm{M}_{\oplus}$ and $100\,\mathrm{M}_{\oplus}$ at $h_0=0.07$ and between $100\,\mathrm{M}_{\oplus}$ and $150\,\mathrm{M}_{\oplus}$ at $h_0=0.08$. As of yet we have been unable to determine a general expression for the critical mass, but assume a proportional relation to the aspect ratio based on our results, which is in agreement with runaway domain studies by \citet{2003ApJ...588..494M}. We do however find that the first rapid migration episode generally sets in when the planet mass (which is kept constant in our simulations) exceeds twice the local thermal mass (see Eq.~\ref{eq:thermalmass}): $M_{\mathrm{p}}\approx 2M_{\mathrm{th}}(r_\mathrm{p})$. Each rapid migration episode is preceded by a buildup phase of slow migration, which delays the onset of rapid migration if this criterion is already met at the initialisation of the simulation (i.e. at $r_\mathrm{p0}$).

The second limiting factor is the emergence of small-scale vortices at the outer gap edge due to the RWI \citep{1999ApJ...513..805L, 2007A&A...471.1043D}. At $\alpha=10^{-5}$ these vortices are no longer effectively suppressed by viscous smoothing, unlike at higher viscosities. Unsuppressed, they diffusively refill the gap and thereby sustain higher migration rates for longer periods in a process known as vortex-assisted migration (\citetalias{2019MNRAS.484..728M}). We find that this causes the planet to stall farther inward than might be expected at this viscosity. Additionally, in some of our simulations, we observe a massive vortex at the outer gap edge breaking the onset of the stall (see $h_0=0.08$ and $h_0=0.09$ in Fig.~\ref{fig:migration_tracks_set1}), a process known as vortex-driven migration \citep{2022A&A...658A..32L}. Vortex-driven migration can even push the planet into the intermittent type-III regime (see $h_0=0.08$ in Fig.~\ref{fig:migration_tracks_set1}). In both cases, the planet eventually comes to a final stall even farther inward.

As such, we find that across most regions of our parameter space the planet comes to its final stall at a radius of $R_\mathrm{stall}\lesssim 15\,\mathrm{au}$, with a potential dust ring no farther out than $R_\mathrm{ring}\lesssim 20\,\mathrm{au}$, which can only account for the innermost rings in ALMA observations. In several cases, but in particular at $\alpha=10^{-3}$, these stalls lead into or occur within the region of the initial gas surface density profile cutoff, close to the inner boundary of the simulation. While we can safely assume that neither the cutoff nor the boundary significantly affects planet migration farther out in the disc (as long as we avoid numerical instabilities such as the one caused by the presence of an inner damping zone; see Appendix~\ref{app:boundary}), this is no longer the case in the cutoff region. Both the tapered gas surface density profile induced by the initial cutoff and the outflow boundary, as well as the proximity to the boundary itself, may modify the torques acting on a planet migrating in this region. Thus, the migration behaviour exhibited by the planet in the cutoff region must be treated with caution. Still, the partial gap-opening observed prior to the planet's arrival at the cutoff indicates that the planet is presumably on track to undergo transition into a type-II stall regardless of boundary effects.

It is only at moderately low viscosity, $\alpha=10^{-4}$, that the planets regularly stall farther outward. The outermost stall we observe is for a $70\,\mathrm{M}_{\oplus}$ planet in a disc with $h_0=0.07$ and $\alpha=10^{-4}$ at $R_\mathrm{stall}\approx 35\,\mathrm{au}$, with a potential dust ring at $R_\mathrm{ring}\approx 50\,\mathrm{au}$. Hence, we conclude that the rings due to a stalled planet can only account for dust rings at small to intermediate stellocentric radii.

\subsubsection{Secondary rings formed by secondary Lindblad spirals}
At low viscosities $\alpha\leq 10^{-4}$, we observe that, as the planet stalls, secondary (and for more massive planets, even tertiary) gaps open inside its orbit. These are created by shocks from the secondary and tertiary Lindblad spirals (\citealp{2018ApJ...859..118B, 2018ApJ...859..119B}, but see also \citealp{2020A&A...637A..50Z}). We find that these structures are more pronounced at higher planet masses. For planets stalling farther inward, they can be difficult to reliably identify due to the close proximity to the simulation boundary. As secondary gaps appear in our simulations only at $R_\mathrm{gap}\lesssim20\,\mathrm{au}$ and share their outer edge pressure maxima with the primary gap, they can account solely for the innermost ALMA rings. They do, however, show that a single planet can still generate multiple rings and gaps once its migration has stalled, as seen previously in \citet{2017ApJ...843..127D}, \citet{2017ApJ...850..201B}, and \citet{2018ApJ...869L..47Z}.

\subsubsection{Remnant rings left by intermittent type-III migration}
When a planet migrates in the intermittent type-III regime, episodes of rapid migration are preceded by slow migration episodes, in which the planet opens a partial gap and builds up inverse vortensity maxima at the gap edges, setting the conditions for the runaway. Once the planet interacts with the material at the inner gap edge in a type-III migration fashion \citep{2003ApJ...588..494M}, it rapidly migrates away, leaving the gap and corresponding ring behind as a remnant \citepalias{2019MNRAS.484..728M, 2020MNRAS.493.5892W}. We have shown in numerous simulations that this mechanism allows for the formation of multiple rings at intermediate radii, $R_\mathrm{ring}\lesssim60\,\mathrm{au}$, by a single planet. Furthermore, our simulation with a $320\,\mathrm{M}_{\oplus}$ planet starting at $r_\mathrm{p0}=200\,\mathrm{au}$ (see Fig.~\ref{fig:migration_tracks_onset320}) illustrates that planets massive enough to enter the intermittent type-III regime at large radii can leave remnant rings far out in the disc -- up to $R_\mathrm{ring}\approx160\,\mathrm{au}$ in this case -- provided their inward migration begins sufficiently far from the star.

However, unlike structures created by the previous two mechanisms, these remnants are no longer directly supported by the planet against viscous diffusion. We find that while remnants diffuse within a few kyr and dissipate entirely on a timescale of $\lesssim100\,\mathrm{kyr}$ at $\alpha=10^{-3}$, they typically persist for the full duration of our simulations at $\alpha\leq10^{-4}$. This corresponds to a structure lifetime of at least $300\text{--}500\,\mathrm{kyr}$ at $\alpha\leq10^{-4}$, a significant fraction of the average disc lifetime \citep{armitage2020}.

\subsection{Constraints on planet and disc properties}
\label{sec:constraints}
Having shown that migrating planets are a viable explanation for multi-ringed disc structures, we now discuss the potential constraints on the physical properties of planets and discs set by the assumption of a planetary origin.

\subsubsection{Conditions for intermittent type-III migration}
As discussed in the previous section, the remnants left by intermittent type-III migration appear to be the primary mechanism by which a migrating planet can form multiple rings and gaps out to large radii. Consequently, if the observed rings are of planetary origin, the conditions for the onset of this migration regime at large radii provide direct constraints on the planet and disc (see also \citealp{2023MNRAS.523.4869W}).

We have shown that, in our models, the first runaway of the intermittent type-III migration typically occurs when the planet mass exceeds twice the local thermal mass (see Eq.~\ref{eq:thermalmass}): $M_{\mathrm{p}}\approx 2M_{\mathrm{th}}(r_\mathrm{p})$. In other words: both the planet mass and disc's aspect-ratio (temperature) profile determine whether and at which radii intermittent migration sets in. We find that for the aspect-ratio profiles investigated in this study, leaving remnants at large radii $r\gtrsim100\,\mathrm{au}$ requires a planet of approximately one Jupiter mass or above.

Another key factor for the occurrence of intermittent type-III migration is the initial gas surface density profile. As all our models adopt the same initial profile, we do not study this dependency in this paper. However, \citet{2003ApJ...588..494M} showed that the occurrence of type-III migration depends on both the disc and planet mass. They found that, at a fixed aspect ratio, the disc must be sufficiently massive to trigger the runaway for a given planet mass, with the runaway domain bounded at both low and high masses by the disc's gravitational-stability limit. More recently, \citetalias{2020MNRAS.493.5892W} further demonstrated that, once type-III migration sets in, it can either be intermittent or smooth, depending on the slope of the initial gas surface density profile, with shallow slopes favouring intermittent behaviour and steep slopes resulting in smooth migration.

\subsubsection{Remnant lifetimes and vortex smear-out}
The remnants of intermittent type-III migration are, however, prone to viscous diffusion. Therefore, low turbulent viscosity, $\alpha\leq 10^{-4}$, is required for the long-term survival of the resulting rings. As discussed in the introduction, this is in accordance with accumulating evidence of low viscosity in protoplanetary discs found in recent years \citep{2016ApJ...816...25P,2018ApJ...869L..46D,2022ApJ...930...11V,2023NewAR..9601674R}.

Unfortunately, low viscosity also tends to lead to non-axisymmetric structures in the form of vortices \citep{2007A&A...471.1043D} rather than the ubiquitous axisymmetric rings seen with ALMA \citep{2018ApJ...869L..41A}. Only in rare cases do the ALMA substructures appear arc-like rather than ring-like (e.g. HD\,143006, MWC\,758). We find, however, that in most cases, the vortices gradually azimuthally smear out to form longer lasting rings. The lower the $\alpha$ value, the longer it takes for the vortices to become ring-like, but the ring-like structures also live proportionally longer. Furthermore, the smear-out timescale tends to be longer farther out in the disc.

In our models with moderately low $\alpha=10^{-4}$, we find that the resulting ring-like structures outlast the smear-out of their progenitor vortex by at least $1.5$ to $5$ (closest to farthest from the star) times their smear-out timescale. For the vortices at the edges of the stalled planet's gap, we typically find smear-out timescales of $\sim50\,\mathrm{kyr}$. The outermost remnant vortices, on the other hand, tend to have smear-out timescales of $\sim200\,\mathrm{kyr}$. However, while our remnant rings show signs of diffusion towards the end of our simulations, they have yet to fully dissolve. Therefore, longer running simulations will be needed in the future to determine the statistics of ring-to-vortex occurrence. Longer running simulations will be especially important at very low $\alpha=10^{-5}$, where vortices only start to smear out towards the end of our simulations and rings show little to no signs of diffusion.

\subsection{Caveats of our methods}
\label{sec:caveats}
In this section, we address important aspects not treated in this paper, be that for the sake of a simple model or to reduce computational demands. We discuss their implications and possible ways to improve.

\subsubsection{Lack of an explicit dust component}
To keep runtimes manageable and avoid introducing additional dust parameters, our models evolved only the gas component. The location and survival of potential dust rings are therefore inferred indirectly from the location, strength, and longevity of gas pressure maxima. While this captures the structures most relevant for dust trapping \citep{1972fpp..conf..211W,2007ApJ...664L..55K}, the omission of an explicit dust component limits what we can infer for both the morphology and persistence of dust rings, as we cannot account for size-dependent drift, diffusion and trapping of dust, or the timescale at which solids actually concentrate at a given pressure maximum. Furthermore, due to the lack of dust-feedback on the gas, the gas profile evolution and planet migration in our simulations may differ from those in a fully coupled gas-dust model (e.g. \citealp{2023MNRAS.523.4869W}). Future work that explicitly evolves a dust population would allow for a more direct assessment of the trapping efficiency and longevity of dust rings. Coupling such models to radiative transfer calculations would further enable a direct comparison with observed continuum structures.

\subsubsection{Self-gravity and the indirect disc-on-disc term}

\begin{figure}
  \centering
  \includegraphics[width=\columnwidth]{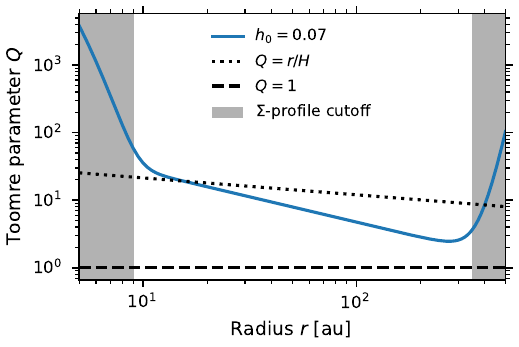}
  \caption{Toomre parameter: Plot of the Toomre parameter, $Q$ (Eq.~\ref{eq:toomre}), at initialisation of our $h_0=0.07$ disc model. For $Q<r/H$, indicated by the dotted line, \citet{2013MNRAS.429..529L} recommended the inclusion of self-gravity when modelling Rossby vortices. The dashed line indicates $Q=1$ as reference for the gravitational stability criterion $Q\gg1$. The grey bands mark the cutoff in the initial gas surface density profile at $r=9\,\mathrm{au}$ and $r=350\,\mathrm{au}$.}
  \label{fig:toomre}
\end{figure}

As the Toomre parameter $Q$ (see Eq.~\ref{eq:toomre}) remains safely above one for all discs considered here -- as illustrated in Fig.~\ref{fig:toomre} for our $h_0=0.07$ model -- the discs are not prone to fragmentation. We therefore omitted disc self-gravity in our simulations.

However, self-gravity studies on the intermittent type-III migration regime by \citetalias{2020MNRAS.493.5892W} showed an effect on planet migration. While their studies indicated that the main features of the migration behaviour remain unchanged, they found that the inclusion of disc self-gravity reduced the initial migration rate, delaying the onset of the first runaway. It also increased the duration of the slow migration episodes between runaways. Therefore, we expect that the inclusion of self-gravity in our models would lead to more pronounced and longer lived remnant structures.

Furthermore, the inclusion of self-gravity can weaken, stretch out, and even suppress vortices \citep{2013MNRAS.429..529L, 2016MNRAS.458.3918Z, 2017MNRAS.471.2204R}. \citet{2013MNRAS.429..529L} in particular recommended the inclusion of self-gravity when $Q<r/H$ in the vicinity of over- and underdensities to accurately model Rossby vortices. As seen in Fig.~\ref{fig:toomre}, most of our $h_0=0.07$ disc has $Q<r/H$ already in its initial state. Thus, self-gravity should be included in future simulations, and may be expected to result in reduced vortex occurrence and smear-out timescales at low viscosities, further shifting the lifetime ratio of ring-like structures to vortices in favour of the former. Overall, this suggests that the inclusion of self-gravity would only lend further credence to the planetary origin of dust rings. Still, the effect of self-gravity on the smooth- and vortex-feedback regimes will need to be further investigated.

We also note that, according to \citet{2025OJAp....8E..84C}, our inclusion of the disc-on-disc indirect term, though not uncommon in recent studies, may not be fully consistent with the omission of disc self-gravity. Even so, the results of the self-gravity studies by \citetalias{2020MNRAS.493.5892W} suggest that this should not affect our main conclusions. Still, it may be related to the instability we observed for simulations with an inner damping zone (see Appendix~\ref{app:boundary}). Therefore, any continuation of our work should incorporate self-gravity where feasible, or otherwise adopt a fully consistent treatment of indirect forces.

\subsubsection{Assumption of a locally isothermal equation of state}
In this work we adopted a locally isothermal equation of state, fixing the temperature as a prescribed function of radius. Numerous studies show, however, that planet--disc interaction is highly sensitive to the efficiency by which the disc can radiatively regulate its temperature.

\citet{2019ApJ...878L...9M,2020ApJ...904..121M,2020ApJ...892...65M} demonstrated that more realistic thermodynamics (such as radiative cooling on the disc surface and even radiation transport in the disc midplane) can substantially alter the dynamics of planet-driven density waves. Thus, the position, morphology, and even number of ring and gap structures may differ from locally isothermal simulations when more realistic thermodynamics are considered. In particular, the locally isothermal approximation is shown to overestimate the planet's gap-opening capabilities, resulting in deeper and wider gaps as well as favouring the formation of secondary gaps, as corroborated by \citet{2020A&A...637A..50Z} and \citet{2024ApJ...961...86Z}.

Thermal physics also affects planetary migration. \citet{2024MNRAS.528.6130Z} found that radiative cooling in inviscid discs induces a baroclinic forcing that weakens the dynamical corotation torque, thereby accelerating inward migration of low-mass planets. Meanwhile, \citet{2025Univ...11....1W} showed that for massive accreting planets, cooling regulates circumplanetary disc asymmetries, which can affect the migration direction.

Taken together, these studies indicate that the locally isothermal approximation can misrepresent both planet-induced disc structures and planetary migration. As a next step, our framework must therefore include at least a $\beta$-cooling prescription, allowing the gas to behave adiabatically while relaxing exponentially towards a prescribed background temperature on a timescale $\beta$. A detailed exploration of how thermal relaxation modifies the intermittent type-III regime is presented in \citet{PAPERII}, where we investigate radiative discs and their observational signatures.

\section{Conclusions}
\label{sec:conclusions}
Our findings indicate that migrating planets could form multi-ringed disc structures similar to those observed with ALMA through three concurrent mechanisms:

\begin{enumerate}
\item Rings form at the gap edge of a stalled (i.e. non-migrating) planet, usually within $r\lesssim 50\,\mathrm{au}$ of the star. We find that stalls occur most reliably and farthest outward at moderately low turbulent viscosity, $\alpha=10^{-4}$.  
\item Secondary rings form at the edge of secondary (or even tertiary) gaps inside the stalled planet's orbit, opened by secondary and tertiary Lindblad spirals \citep{2018ApJ...859..118B, 2018ApJ...859..119B} at low viscosity, $\alpha\leq10^{-4}$. These are typically located within $r\lesssim 20\,\mathrm{au}$.
\item Remnant rings form at the edge of partial gaps outside the planet's orbit. These are built up in slow migration phases and are then left behind in type-III runaway migration episodes occurring in sufficiently massive discs \citepalias{2019MNRAS.484..728M, 2020MNRAS.493.5892W}. For this behaviour to emerge prior to a stall, a critical planet mass that is dependent on the disc aspect ratio ($M_{\mathrm{p}}\gtrsim100\,\mathrm{M}_{\oplus}$ in our simulations) is required. The first runaway then typically sets in at twice the thermal mass: $M_{\mathrm{p}}\approx 2\,M_{\mathrm{th}}(r_\mathrm{p})$. Thus, more massive planets ($M_{\mathrm{p}}\sim320\,\mathrm{M}_{\oplus}$ in our simulations) can leave rings even at large stellocentric radii, out to at least $r\sim 150\,\mathrm{au}$. Remnants dissipate quickly through viscous diffusion at $\alpha=10^{-3}$, but persist for at least $300\text{--}500\,\mathrm{kyr}$ at $\alpha\leq10^{-4}$, a significant fraction of an average disc lifetime \citep{armitage2020}.
\end{enumerate}

At $\alpha\leq10^{-4}$, the dust-trapping pressure maxima located at the gap edges start out as large-scale vortices, which gradually smear out into ring-like structures. Both the smear-out and remnant-dissipation timescales increase with lower $\alpha$. The smear-out timescale also increases farther out in the disc. We find that ring-like structures outlast their progenitor vortices by at least $1.5$ to $5$ times the smear-out timescale at $\alpha=10^{-4}$. This should result in ring-like structures being more prevalent than asymmetries in observations of protoplanetary discs, in agreement with the general trend in ALMA observations \citep{2018ApJ...869L..41A}. Longer simulations will be required to compare the statistics of ring-to-vortex occurrence to observations, especially at lower viscosities ($\alpha<10^{-4}$) and larger radii.

We conclude that the planetary origin hypothesis of multi-ringed structures out to $r\sim 150\,\mathrm{au}$ is viable for planets of approximately one Jupiter mass or above in discs with low viscosity, $\alpha\lesssim10^{-4}$, that are sufficiently massive to enable type-III migration.

\section*{Data availability}
The data obtained from our numerical models are available from the corresponding author upon reasonable request.

\begin{acknowledgements}
  We thank Andrej Herrmann, whose earlier simulations inspired parts of this work. We also thank Alexandros Ziampras and Philippine Griveaud for their helpful advice and insightful discussions. We acknowledge support by the High Performance and Cloud Computing Group at the Zentrum für Datenverarbeitung of the University of Tübingen, the state of Baden-Württemberg through bwHPC and the German Research Foundation (DFG) through grant INST 37/935-1 FUGG. K.M.W. acknowledges funding from the European Union under the European Union's Horizon Europe Research and Innovation Programme 101124282 (EARLYBIRD). Views and opinions expressed are, however, those of the authors only and do not necessarily reflect those of the European Union or the European Research Council. Neither the European Union nor the granting authority can be held responsible for them. T.R. and C.P.D. acknowledge funding from the Deutsche Forschungsgemeinschaft (DFG) research group FOR 2634 (grants KL 650/29-2 + 30-2 and DU 414/22-2 + 23-2) “Planet Formation Witnesses and Probes: Transition Disks”. All plots in this paper were made with the \texttt{python} package \texttt{matplotlib} \citep{matplotlib}.
\end{acknowledgements}

\bibliographystyle{aa}
\bibliography{aa60768-26}

\begin{appendix}
\nolinenumbers
\section{Resolution convergence study}
\label{app:resolution}
\begin{figure}
  \centering
  \includegraphics[width=\columnwidth]{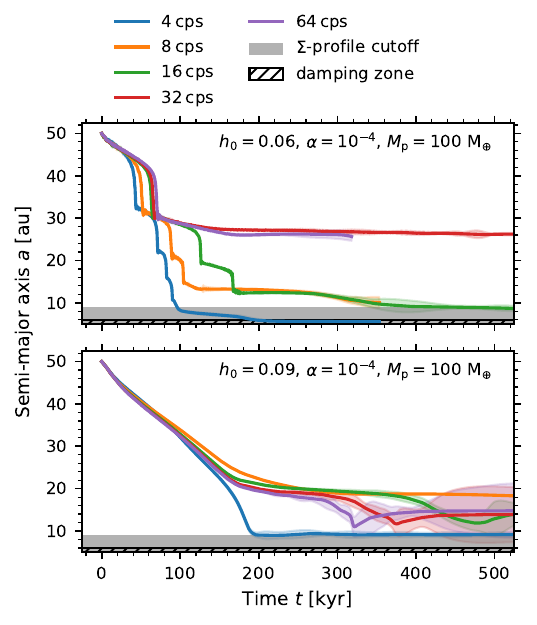}
  \caption{Migration tracks of the resolution convergence study: Time evolution of the semi-major axis, $a$, of a $100\,\mathrm{M}_{\oplus}$ planet for two values of the disc aspect ratio, $h_0$, at $\alpha=10^{-4}$ and various grid resolutions. The presence and extent of a pale coloured region around a migration track indicates whether and to what degree the planet's eccentricity is excited, as these regions span the extent between the planet's apsides. The grey band marks the cutoff in the initial gas surface density profile at $r=9\,\mathrm{au}$. The black hatched area marks the damping zone at $r\leq6\,\mathrm{au}$.}
  \label{fig:resolution_study}
\end{figure}

Figure~\ref{fig:resolution_study} presents the resolution convergence study mentioned in Sect.~\ref{sec:setup}. As these were among our earliest simulations, they still feature an inner damping zone at $r\leq6\,\mathrm{au}$, which triggers an instability in the $h_0=0.09$ disc. This instability results in the excited planetary eccentricity in the later stages of migration, $t\gtrsim300\,\mathrm{kyr}$. For further details on this instability, see Appendix~\ref{app:boundary}.

For the $h_0=0.06$ model, we find that the planet migrates in the intermittent type-III regime across all tested resolutions. With increasing resolution, the number of type-III runaway episodes decreases, while the duration of the preceding slow migration episodes increases. For a given runaway episode (e.g. the first, second, third, etc.), the stellocentric distance covered is nearly resolution-independent. As a result, the transition to type-II migration occurs farther out at higher resolutions. While simulations at lower resolutions show various divergences and eventually migrate into the $\Sigma$-profile cutoff and towards the inner boundary, the $32$\,cps and $64$\,cps simulations largely converge, each featuring only a single runaway episode followed by a type-II stall at $r_\mathrm{stall}\approx26\,\mathrm{au}$.

For the $h_0=0.09$ model, on the other hand, we find that the planet migrates in the smooth- or vortex-feedback regime across all tested resolutions. All resolutions show broadly convergent migration behaviour until the disc becomes unstable at late times, $t\gtrsim300\,\mathrm{kyr}$, except for the $4$\,cps simulation, which diverges as the others transition to type-II migration.

\section{Inner boundary treatment: Effect on migration and avoiding numerical instabilities}
\label{app:boundary}
\begin{figure}
  \centering
  \includegraphics[width=\columnwidth]{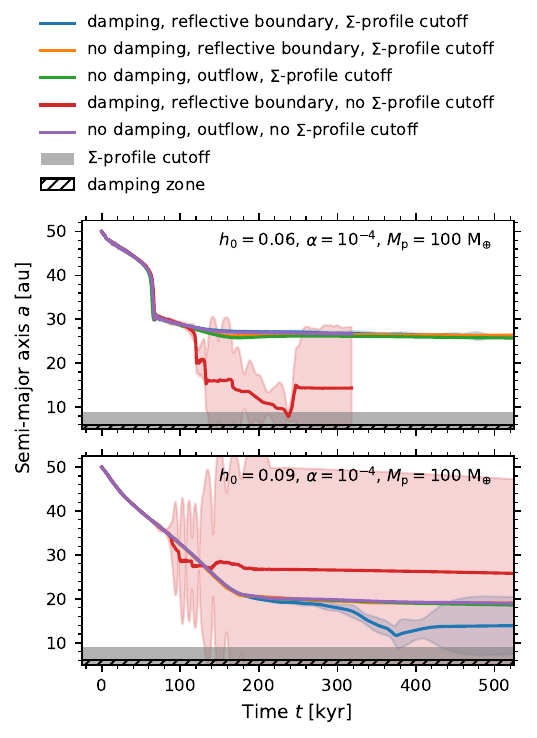}
  \caption{Same as Fig.~\ref{fig:resolution_study} but for various inner boundary treatments at a grid resolution of $32\,\mathrm{cps}$.}
  \label{fig:boundary_study}
\end{figure}

Figure~\ref{fig:boundary_study} illustrates how different inner boundary treatments affect the planet's migration behaviour and the instability observed in our earliest simulations (see Appendix~\ref{app:resolution}). We test the effect of the inner damping zone at $r\leq6\,\mathrm{au}$ and the inner initial gas surface density profile cutoff at $r=9\,\mathrm{au}$. To avoid possible interference from reflections or pileups in the absence of the damping zone, these simulations employ an outflow condition at the inner boundary. To verify that this precaution does not bias our results, we also include a simulation without the inner damping zone, but with reflective boundary conditions and the cutoff.

We identify the inner damping zone as the primary origin of the instability, as only simulations that include it exhibit the strongly excited planetary eccentricity characteristic of the unstable behaviour. The initial gas surface density profile cutoff can suppress this instability almost entirely in the $h_0=0.06$ model and partially mitigate it in the $h_0=0.09$ model. In contrast, all simulations without the inner damping zone remain entirely free of the instability, while neither the cutoff nor the inner boundary condition (reflective or outflow) significantly affects planetary migration, at least for $r\gtrsim18\,\mathrm{au}$.

Considering these results, we interpret the eccentricity growth in the presence of the inner damping zone as a numerical instability. It can be safely avoided by removing the damping zone. In its absence, an outflow condition at the boundary can be used to prevent interference without introducing any bias.

\onecolumn
\section{Further azimuthal average overviews}
\label{app:aziavg}
\begin{figure*}[ht!]
  \centering
  \includegraphics[width=\textwidth]{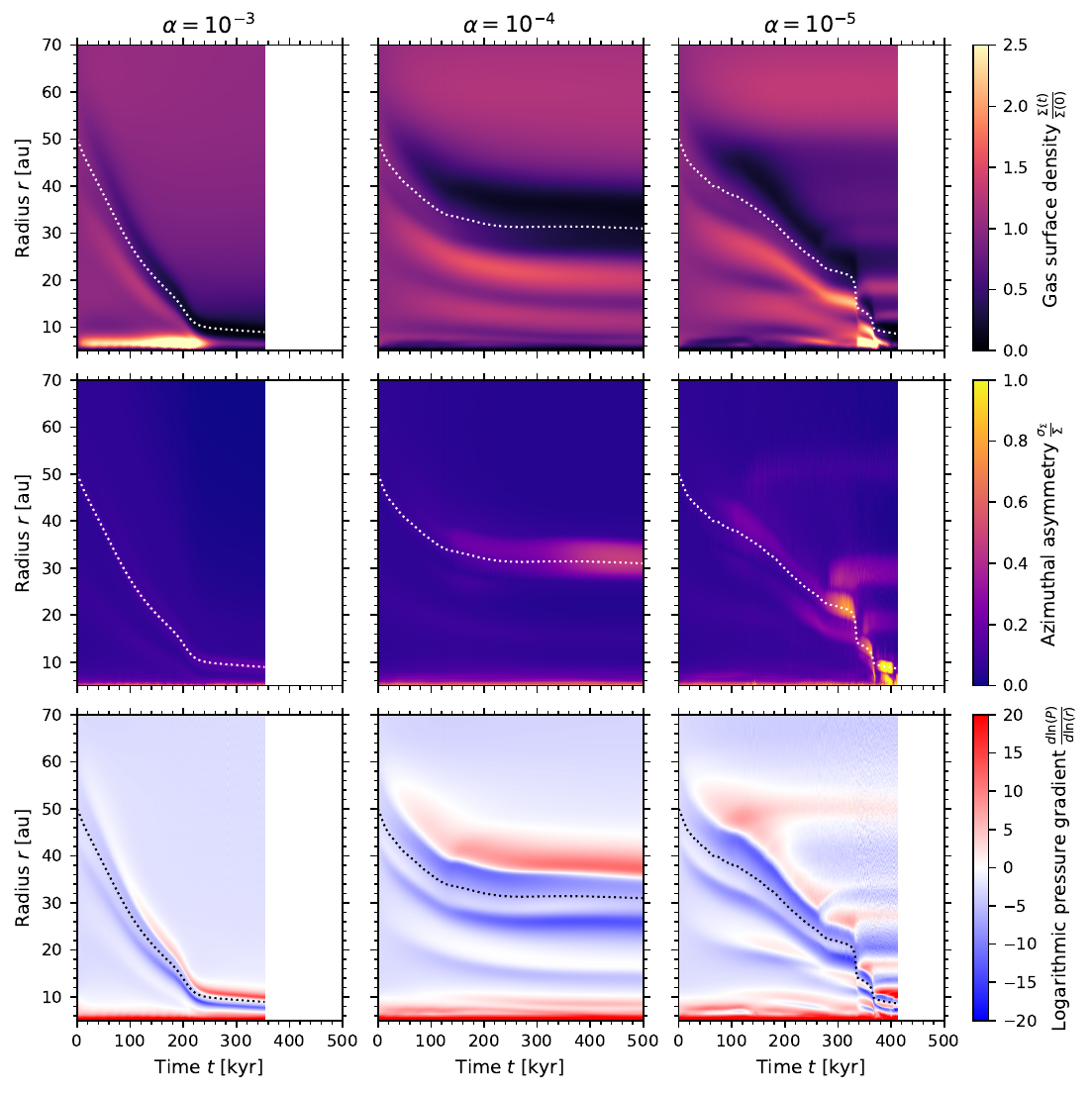}
  \caption{Azimuthally averaged time evolution of the gas for a $100\,\mathrm{M}_{\oplus}$ planet in a disc with $h_0=0.08$ at different $\alpha$: The dotted line represents the planet's migration track. \textit{Top row}: Azimuthally averaged gas surface density, $\Sigma$, normalised to initial values. \textit{Bottom row}: Logarithmic gas pressure gradient at the midplane. Stark white regions at the transition from red to blue (towards larger $r$) are indicative of local gas pressure maxima, likely candidates for dust traps. \textit{Middle row}: Azimuthal standard deviation of $\Sigma$ normalised to its current value at time, $t$. This illustrates the asymmetries in the gas surface density, which are indicative of vortices. Note: this quantity is only meaningful outside deep gaps.}
  \label{fig:set1_h008_aziavg}
\end{figure*}

\begin{figure*}[ht!]
  \centering
  \includegraphics[width=\textwidth]{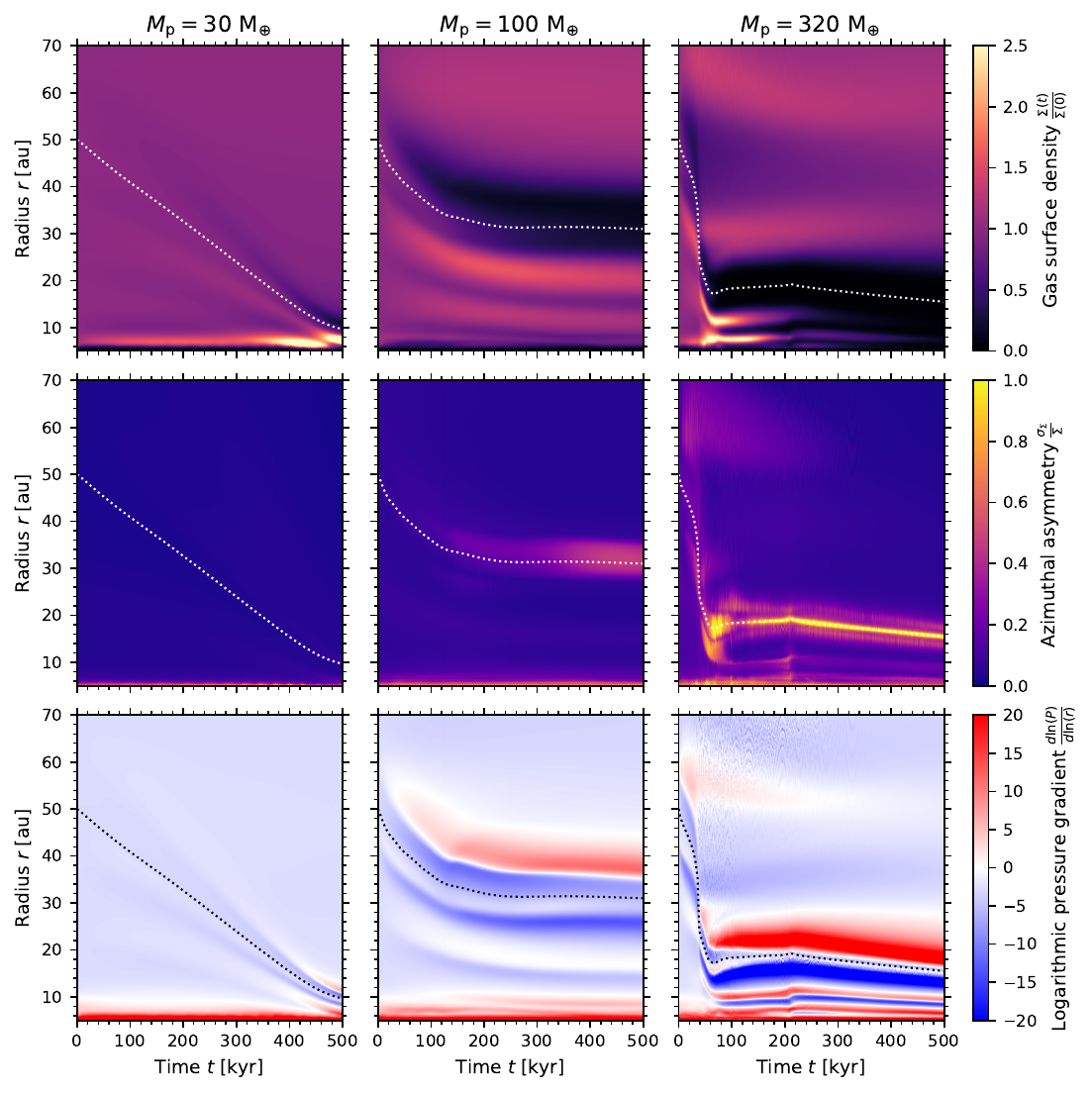}
  \caption{Same as Fig.~\ref{fig:set1_h008_aziavg} but for various planet masses in a disc with $h_0=0.08$ at $\alpha=10^{-4}$. For comparison, the middle columns in this figure and Fig.~\ref{fig:set1_h008_aziavg} depict the same simulation.}
  \label{fig:set2_h008_aziavg}
\end{figure*}

\clearpage

\section{Full second set of simulations (effect of the planet mass)}
\label{app:set2}
\begin{figure*}[ht!]
  \centering
  \includegraphics[width=\textwidth]{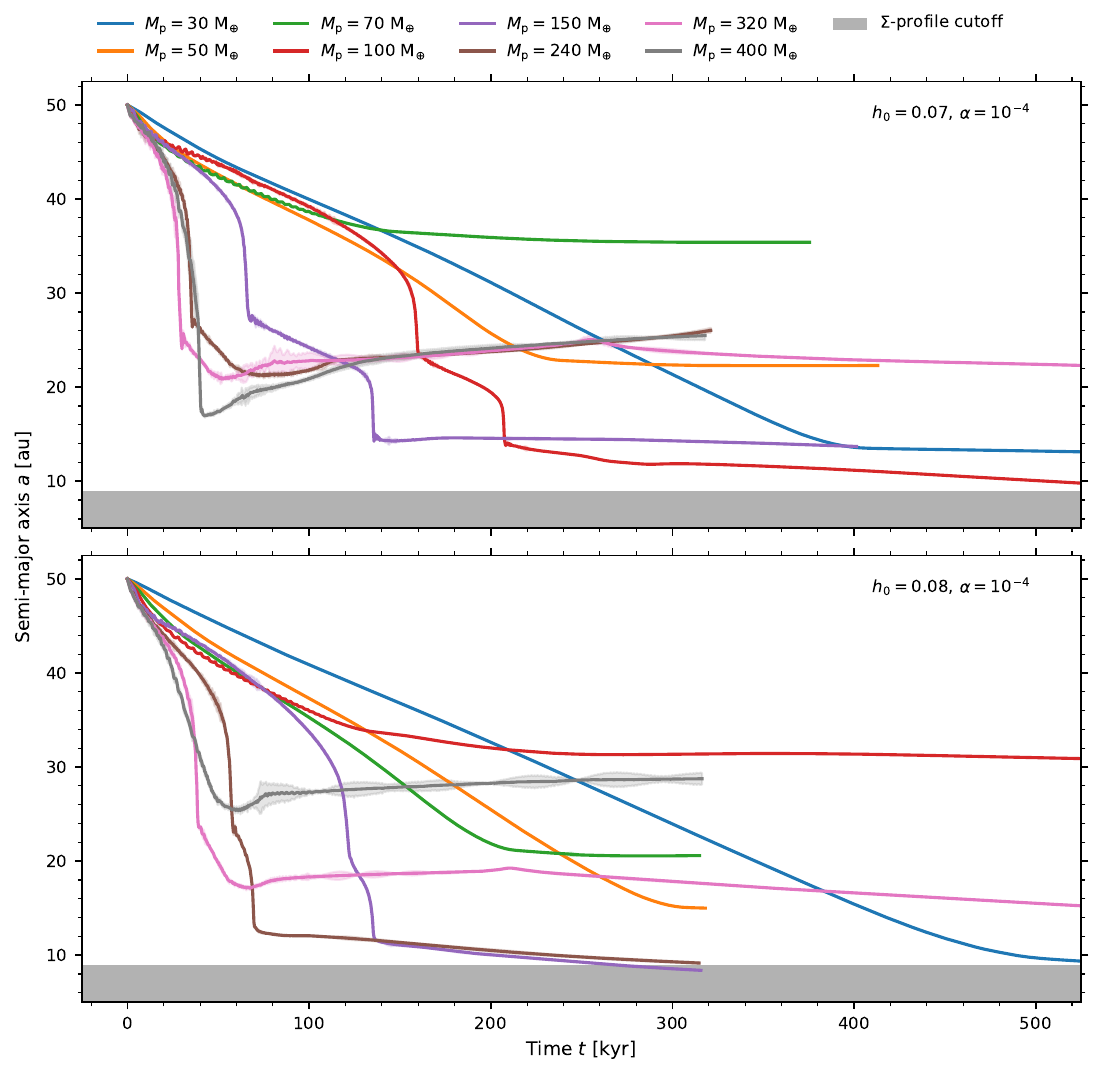}
  \caption{Migration tracks of the second set of simulations: Time evolution of the semi-major axis, $a$, of planets with masses between $30\,\mathrm{M}_{\oplus}$ and $400\,\mathrm{M}_{\oplus}$ for two values of the disc aspect ratio, $h_0$, at $\alpha=10^{-4}$. The presence and extent of a pale coloured region around a migration track indicates whether and to what degree the planet's eccentricity is excited, as these regions span the extent between the planet's apsides. The grey band marks the cutoff in the initial gas surface density profile at $r=9\,\mathrm{au}$.}
  \label{fig:migration_tracks_set2}
\end{figure*}

\end{appendix}

\end{document}